\documentclass[a4paper,fleqn]{cas-dc}

\usepackage[numbers]{natbib}
\usepackage{amssymb}
\usepackage{amsmath}
\usepackage{pifont}
\newcommand{\cmark}{\ding{51}} 
\newcommand{\xmark}{\ding{55}} 
\newcommand{\cxmark}{\ding{55}}
\usepackage{makecell}
\usepackage{longtable}
\usepackage{soul,color}
\usepackage{graphicx}
\usepackage{textcomp}
\usepackage{xcolor}
\usepackage{float}

\def\tsc#1{\csdef{#1}{\textsc{\lowercase{#1}}\xspace}}
\tsc{WGM}
\tsc{QE}
\tsc{EP}
\tsc{PMS}
\tsc{BEC}
\tsc{DE}

\begin{document}
\let\WriteBookmarks\relax
\def\floatpagepagefraction{1}
\def\textpagefraction{.001}
\shorttitle{High Altitude Platforms}
\shortauthors{S. Çoğay et~al.}

\title [mode = title]{HAP Networks for the Future: Applications in Sensing, Computing, and Communication}                      
\tnotemark[1]

\tnotetext[1]{This work was supported in part by Istanbul Techni-
cal University, Department of Scientific Research Projects
(ITU-BAP) under Grant 45375; and in part by Türkiye-
South Korea Bilateral Call under Project TUBITAK 2523-
123N821; and in part by the National Research Foundation
of Korea under Grant RS-2023-NR121113.}


\author[1]{Sultan Çoğay}[
                        orcid=0000-0002-4707-6032]


\affiliation[1]{organization={Department of Computer Engineering,Istanbul Technical University},
                city={Istanbul}, 
                country={Turkey}}

\author[1]{T.Tolga Sarı}[orcid=0000-0002-2100-4890]

\author[2]{Muhammad Nadeem Ali}[
   orcid=0000-0002-1240-8148]


\affiliation[2]{organization={Department of Software and Communications Engineering, Hongik University},
                postcode={30016}, 
                city={Sejong},
                country={South Korea}}

\author[2]{Byung-Seo Kim}[orcid=0000-0001-9824-1950]

\affiliation[3]{organization = {BTS - Digital Twin Application and Research Center, Istanbul Technical University},
city={Istanbul},
country={Turkey}}

\author[1,3]{Gökhan Seçinti}[orcid=0000-0003-0640-8368]
\cortext[cor1]{Corresponding author: Sultan Çoğay(cogay@itu.edu.tr)}


\begin{abstract}
High Altitude Platforms (HAPs) are a major advancement in non-terrestrial networks, offering broad coverage and unique capabilities. They form a vital link between satellite systems and terrestrial networks and play a key role in next-generation communication technologies. This study reviews HAP network applications, focusing on advanced airborne communications, integrated sensing, and airborne informatics. Our survey assesses the current state of HAP-centric applications by examining data processing, network performance, computational and storage requirements, economic feasibility, and regulatory challenges. The analysis highlights the evolving role of HAPs in global communication and identifies future research directions to support their deployment.
\end{abstract}




\begin{keywords}
High Altitude Platforms (HAPs)\sep  Non-terrestrial Networks \sep Space-air-ground Integrated Network \sep Aerial networks \sep 6G
\end{keywords}

\maketitle

\section{Introduction}
Rapid advances in communication and network technologies have accelerated the use and development of non-terrestrial networks (NTNs). As global connectivity demands grow, the use of NTNs is becoming increasingly widespread. NTNs are an alternative for extending coverage to underserved, rural, and remote areas\cite{azari2022evolution}. NTN integration enhances the capabilities of existing networks. These developments are important for a wide variety of application scenarios across multiple sectors. In this way, NTNs significantly expand the reach and flexibility of existing communication infrastructures\cite{wang2024unmanned}. The adoption of NTNs is a significant advancement in terms of expanding communication coverage.

\begin{figure}
    \centering
    \includegraphics[width=.95\linewidth]{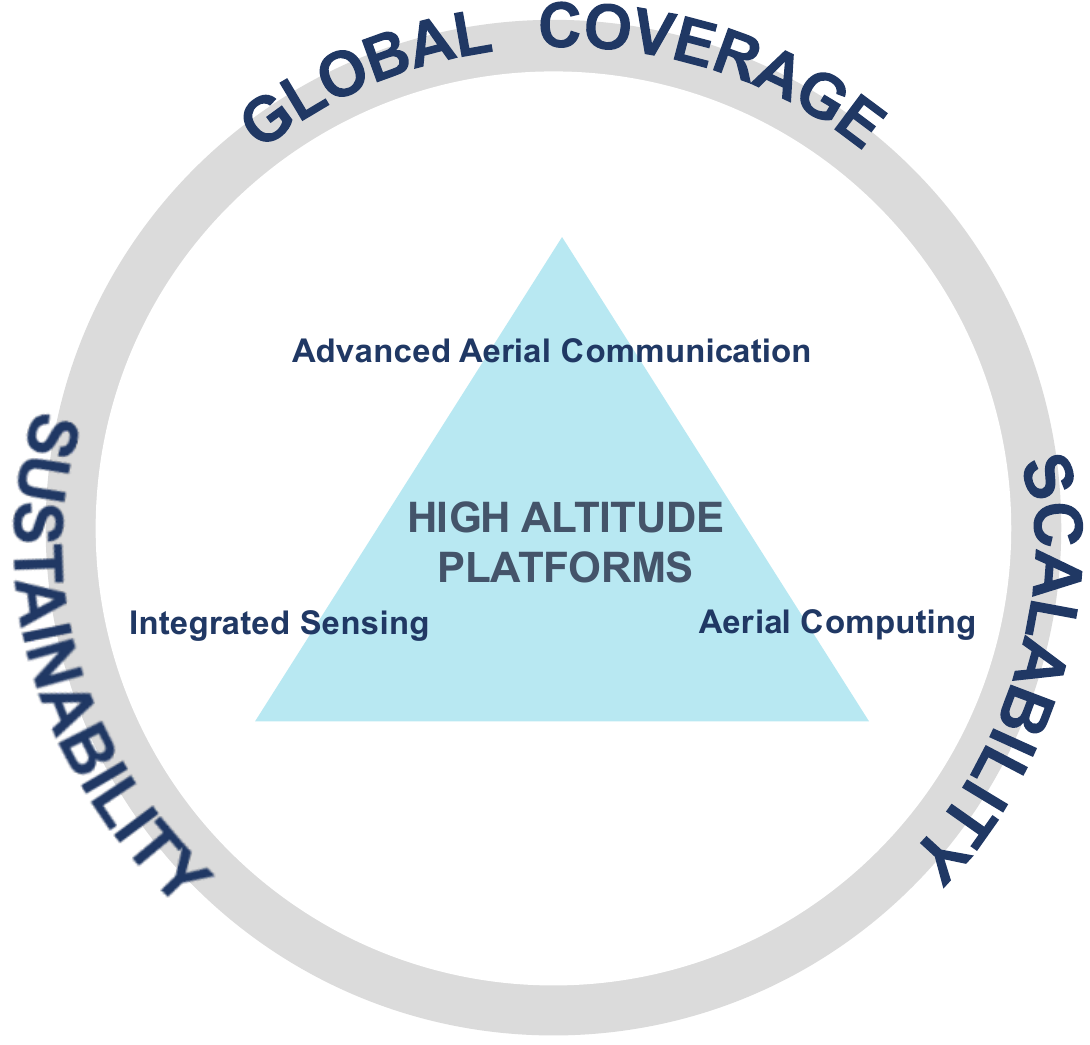}
    \caption{The categorization of the HAP studies in the survey.}
    \label{fig:fig1}
\end{figure}

Technical, satellite and terrestrial networks constitute the basis of global connectivity. However, geographical and infrastructural limitations require different technologies. HAPs and aerial communication technologies such as unmanned aerial vehicles (UAVs) are incorporated into these systems to meet these needs\cite{bakambekova2024interplay}. Thanks to their technical characteristics, HAPs can achieve wide coverage and stable connectivity over large geographical areas. To provide these connections, they combine satellite networks with terrestrial infrastructure. Therefore, HAP networks offer flexible, affordable, and rapidly deployable solutions for many uses, address challenging issues such as communication and computing resource allocation\cite{kurt2021vision}. Additionally, as shown in Figure \ref{fig:fig1}, HAPs can provide global coverage, scalability, and sustainability for next-generation communication and networking applications. Because of their long mission times and extensive coverage areas, these platforms also provide a cheap alternative, especially for thinly populated areas.

HAP networks are mostly strong in their capacity to enable essential communication links for people, communities, emergency response teams, and different industry players all around. By including cutting-edge sensing technologies into HAP systems, their capacity to provide integrated sensing and communication is further improved, so facilitating real-time data collecting, dynamic situational awareness, and quick responsiveness across fields including agriculture, environmental monitoring, disaster management, and intelligent transportation systems\cite{nauman2023empowering}. This then facilitates more efficient use of resources and supports proactive decision-making practices.

By means of constant monitoring of crop health, soil conditions, and climatic parameters, HAPs provide precision management features in the agricultural sector, so greatly enhancing resource efficiency and productivity. In the maritime industry, HAPs enhance navigation safety, enable robust maritime communication, and extensive monitoring to effectively manage maritime traffic. By employing HAP networks, it is possible to create smarter, safer, and more sustainable metropolitan areas\cite{nauman2023empowering,belmekki2024cellular}. Within the scope of smart cities, applications such as traffic control, environmental monitoring, and public safety can also be further developed.

Research and development efforts intended to facilitate the integration of HAPs into next-generation networks are ongoing by authorities and researchers. Developments in areas such as spectrum access to increase the use of HAPs have provided solutions to altitude-related problems\cite{belmekki2024cellular}. This has largely paved the way for HAPs to become an essential element for heterogeneous networks.

\begin{figure}
    \centering
    \includegraphics[width=.75\linewidth]{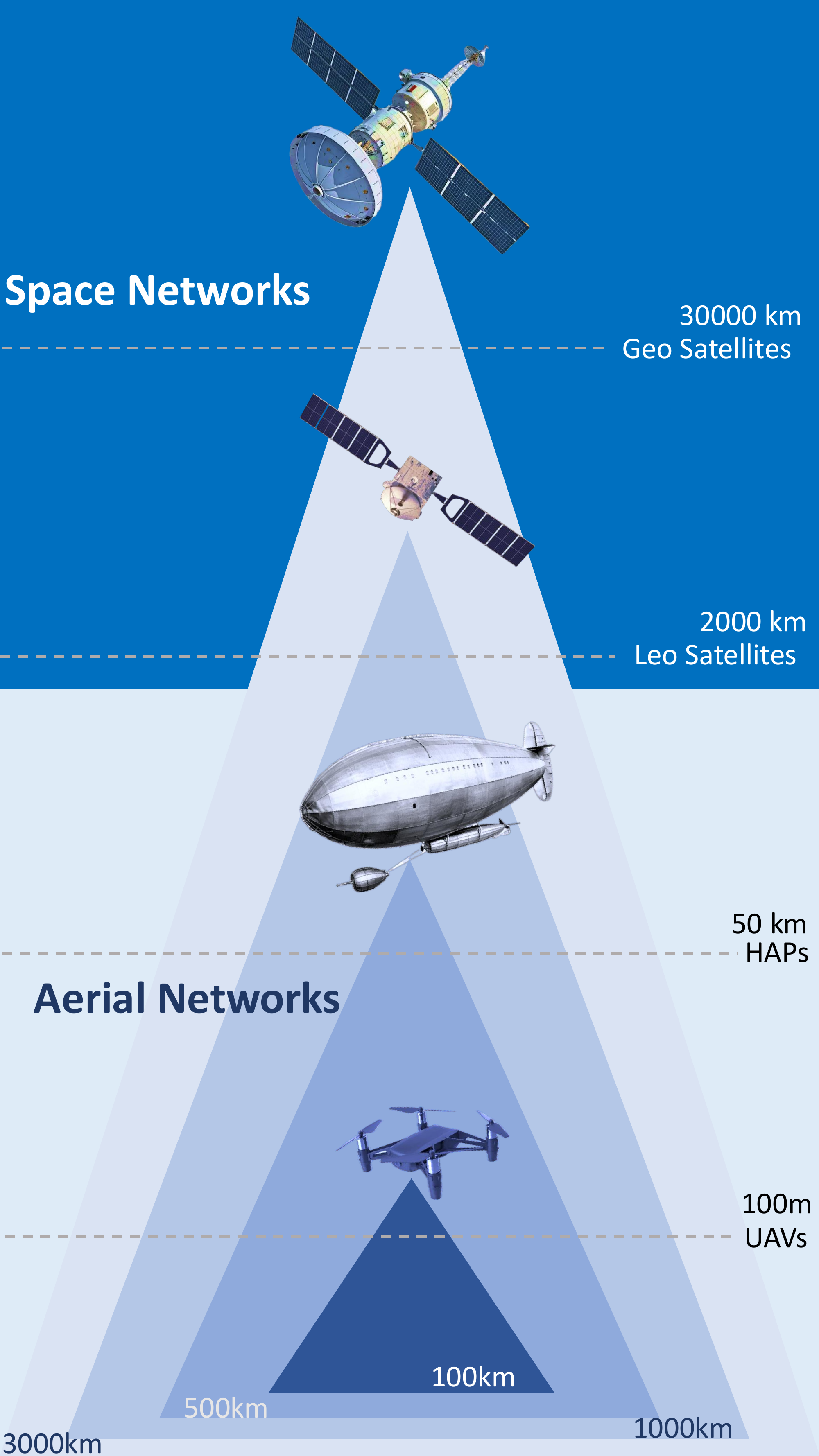}
    \caption{Altitude and Coverage of Space and Aerial Networks}
    \label{fig:intro2}
\end{figure}

HAPs can be integrated into space and air-based network systems, significantly increasing the capacity and efficiency of communications infrastructure. HAPs are platforms typically positioned at altitudes of 20 to 50 kilometers and, when used in coordination with other air and space assets such as satellites and UAVs, can provide continuous, high-quality communications services over large geographical areas as shown in Figure \ref{fig:intro2}. One advantage of HAPs is that they offer lower latency and higher bandwidth than conventional satellite systems. They also complement space-based communication systems by reducing signal attenuation and interruptions due to atmospheric conditions. HAPs can strengthen connectivity between air and space networks, especially by integrating with LEO (Low Earth Orbit) and GEO (Geostationary Earth Orbit) satellite networks~\cite{kakati2025towards}. Among several benefits over conventional satellite systems, HAP networks offer much lower latency, cost economy, and more operational freedom. With modern payloads and solar-powered propulsion systems, HAPs can keep running continuously, so providing strategic resource management tools, consistent communication, and robust computing support. Moreover, when integrated with agile UAV systems acting as mobile edge relays or servers, HAPs enable dynamic, localized data processing and targeted service delivery; these are basic needs for low-latency, high-reliable applications~\cite{kurt2021communication,9887916}. Particularly in cases requiring fast and heavy data traffic (e.g., major events or emergency operations), this synergy is quite crucial.

The introduction of emerging technologies, such as Reconfigurable Intelligent Surfaces (RIS), further enhances the adjustable capabilities of aerial communication networks~\cite{umer2025reconfigurable,gao2021aerial}. Dynamic handling of electromagnetic signal paths by RIS-equipped HAPs provides countermeasures against environmental-related interference and overcomes non-line-of-sight (NLoS) propagation issues. The result is a significant increase in spectral efficiency and improved communication robustness. Artificial Intelligence (AI)-driven optimization and algorithms empower intelligent resource management, adaptive beamforming, and automatic network reconfiguration that creates self-optimizing and flexible communication ecosystems~\cite{shamsabadi2025interference}. 

Conventional centralized cloud computing infrastructures are proving inadequate to support real-time, low-latency, and context-aware computing services which leads to the growing adoption of edge-based computing frameworks. The combination of HAP and Unmanned Aerial Vehicle (UAV) technologies into Mobile Edge Computing (MEC) systems increases the provision of computing services in areas with limited terrestrial infrastructure. Hybrid systems improve communication latency, bandwidth, energy consumption, and user-specific Quality of Service (QoS) requirements. As a result, data can be processed closer to its source, supporting more responsive applications and real-time decision-making processes~\cite{lakew2021intelligent}.

Aerial technologies also provide advanced edge capabilities such as intelligent content caching and FL to significantly reduce latency, preserve bandwidth, and preserve data privacy using localized data processing. Acting as federated aggregators, HAPs efficiently coordinate distributed AI training between ground devices and UAVs\cite{10250790}. This method gives a significant advantage, particularly for applications that require sensitive data to remain local, and eliminates the need to copy large datasets to a central server.

Even though there exists a lot of development, there are still many problems to solve. Important areas of research include managing dynamic network topologies in real time, optimizing aerial trajectories that are limited by energy, and coordinating multiple agents well. Regulatory factors, like how frequencies are shared around the world, how aerial platforms work across borders, and data privacy standards, will affect how quickly HAP networks are adopted and integrated all over the world. Also, ensuring that systems from different manufacturers can work together and keeping these systems safe from cyberattacks are two important issues~\cite{saber2024security}.

Still, the great promise and quick technological developments point to a bright future for HAP-UAV-enabled edge computing and NTNs, so rendering them absolutely essential components of next-generation wireless communication networks~\cite{qiu2019air}. The limits of global connectivity will be redrawn as these technologies develop, knowledge will be democratized, and new avenues for invention will open out over many different fields. This metamorphosis is a pillar of a more linked, intelligent, and resilient world vision as much as a technological development. For all these reasons, we propose a comprehensive survey about the newest application areas of HAP networking, including three primary function capabilities: advanced aerial communication \& networking, integrated sensing, and advanced aerial computing.
\begin{figure*}
    \centering
    \includegraphics[width=0.99\linewidth]{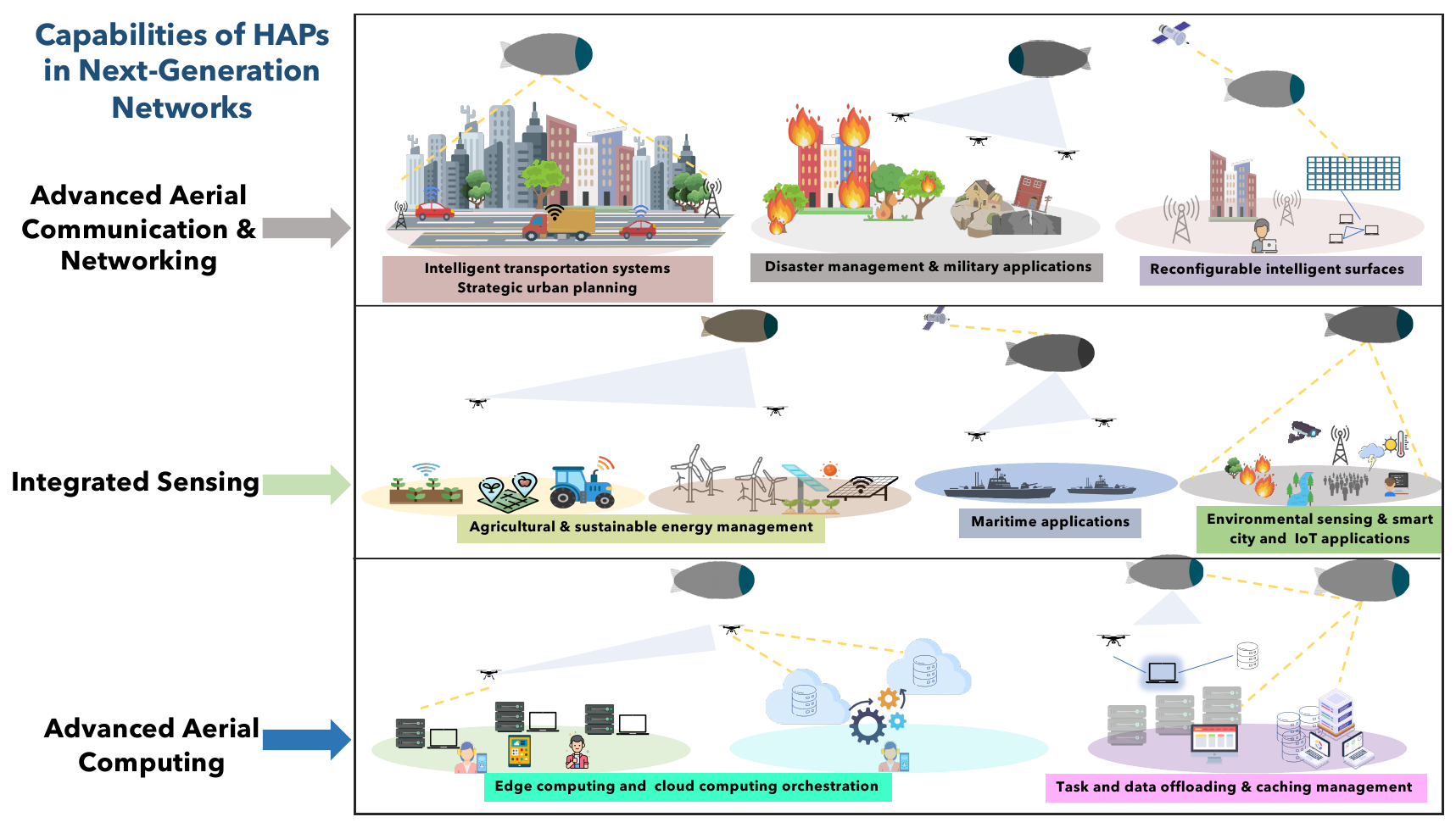}
    \caption{Taxonomy and organization of the survey}
    \label{figure:hap1}
\end{figure*}

Our aim with the survey is to systematically identify and examine the trends, advantages, and disadvantages across different HAP applications. The survey is thoroughly designed to take into account numerous important factors that impact HAP applications.  These factors include data properties such as the data types being processed, the volume of data handled, along with the specific requirements for realtime or batch processing. Furthermore, the investigation evaluates network performance by examining essential metrics such as bandwidth, latency and coverage that HAP networks can support. It also examines at how much processing power and storage space these systems need to run. We discuss economic aspects like cost-benefit and financial feasibility, as well as the geographical focus of different HAP uses. The survey additionally investigates at sector orientation by looking at how industries are using HAP technology and the rules that govern its use, such as privacy rules and licensing requirements. The social effects of HAP interventions, such as improving communication and connectivity in areas that lack them, are also very important.
In the literature, there are surveys investigating studies related to HAP. While~\cite{kurt2021vision,dicandia2022space,riccio2025comprehensive,saeed2021point,alam2023high}, discuss communication-focused and technical design approaches, our review is detailed in several aspects.
Our survey gives a comprehensive overview and application perspective on the potential of HAPs, delivering a notable advantage. This advantage derives from its extensive and balanced classification, which bases HAPS applications on three main and balanced functions: Advanced Air Communication \& Network Deployment, Integrated Sensing, and Advanced Air Computing. Furthermore, the survey does not limit its evaluation of HAP-based systems to technical metrics.  The survey demonstrates the role of HAPs in a wide range of network layer and application layer protocols, including signal manipulation with RIS, data privacy and AI applications with FL, and low latency with MEC. Finally, it clearly highlights HAPS's regulatory challenges, such as international frequency allocation and data privacy standards, as well as its legal and social impacts, such as its potential to create social impact in areas without connectivity.

Following that,  comparison tables systematically organize these dimensions across different HAP applications to facilitate a thorough analysis. Tables have columns for specific metrics, which allows the comparison of how different HAP technologies work with each other in a detailed and organized way. The information gathered from this survey will show what HAP technologies can do now and which need further research. This will help HAP technologies get better and be used more widely. As a result, the main contributions are listed below: 
\vspace{0.5cm}

\begin{itemize}
    \item We propose a detailed survey of the latest applications of HAP networks, focusing on three critical capabilities: advanced aerial networking and communication, integrated sensing, and advanced aerial computing.
    \item The survey provides a comprehensive comparison tables for assessing HAP applications using  communication, networking, and computing metrics.
    \item The study methodically identifies and examines trends in implementing HAP technologies in several industries (such as telecommunications, energy sector, military, smart city and transportation etc.) emphasizing strengths and areas that need improvement.
    \item For each subcategory in the survey, a comparative analysis and detailed comparison between studies is provided.

\end{itemize}
\vspace{0.3cm}
In the rest of the paper, we present the preliminary study by three main tables in subsections. Section \ref{section1} and Table \ref{tab:HAP-ITS}  provides research on advanced aerial communication and networking, including ITS, IRS, and applications in disaster contexts. 
Section \ref{section2} and Table \ref{tab:is} presents research focused on aerial integrated sensing, covering monitoring, and positioning systems. Section \ref{section3} and Table \ref{tab:aac} presents a summary of studies that examine cloud-based, edge-based, task and data offloading, and caching applications in the context of advanced aerial computing. 

\section{HAP-Centric Applications}
To focus on the unique characteristics of HAP-centric applications across various domains, this section is divided into three subsections as shown in Figure \ref{figure:hap1}. Each subsection provides a detailed explanation and is accompanied by tables summarizing relevant research and applications.

\subsection{Advanced Aerial Communication \& Networking} \label{section1}
HAPs offer a promising solution to extend radio coverage to unserved or underserved areas. In addition to facilitating broadband connectivity over large distances through high-altitude operations, HAPs can provide uninterrupted communications in hotspot areas or disaster-affected areas.

Academia has explored the capability of collaborative systems that integrate HAPs with satellites and UAVs to further explore airborne communications and networking. These systems can provide complementary benefits, such as persistent connectivity from HAPs, flexible UAV deployment, and wide-area coverage from satellites. 

The numerous uses for HAP technology illustrate its adjustability.  HAPs have the potential to alter data transfer and communication in a variety of applications. Additionally, their inclusion in complex network architectures can greatly improve performance and connectivity under difficult conditions.


\subsubsection{Intelligent Transportation Systems}

The primary purpose of intelligent transportation systems (ITS) is to ensure that vehicles are aware of their surroundings. Transportation systems can also provide solutions to “blind spot” issues that can lead to accidents and inefficiency. As the number of vehicles and demand for transportation increase, traffic problems also increase. Solving increasing traffic problems increases the need and demand for modern transportation systems and traffic management solutions.

As innovative mobility solutions rapidly grow, ITS applications are becoming more widespread. Consequently, they assist in fleet optimization, traffic congestion control, accident analysis, route planning, and the development of predictive traffic models. HAPs are used as innovative technologies that greatly improve ITS efficiency and capabilities. HAPs are technologies that can provide wide coverage, continuous connectivity, and computing power to support ITS operations. Furthermore, unlike satellite systems with high latency or terrestrial cellular networks with infrastructure restrictions, they provide low-latency, high-capacity communication services over long distances. Supporting ITS in urban environments, intercontinental routes, and isolated areas where usual infrastructure is limited or non-existent makes these systems highly beneficial.
Various studies exist to address infrastructural challenges.
~\cite{popoola2022exploiting} specifically addresses the problem of coverage limitations in traditional infrastructure like Roadside Units (RSUs).  The study addresses the challenges of providing reliable, high-data-rate vehicle communication services in rural areas where existing technologies such as DSRC offer insufficient bandwidth. To overcome these issues, the research proposes using a HAP operating in the millimeter wave (mmWave) frequency band equipped with a large antenna array.
~\cite{djihane2024exploring},  focuses on improving connectivity, coverage, and reliability, including UAVs, HAPS, and satellites. Thus, they offer solutions for how next-generation ITS can be developed. 
For NTN-based ITS, technologies like Free Space Optics (FSO) communication and millimeter wave links offer different solutions. Despite atmospheric attenuation, high performance can be achieved with these solutions.

One of the most significant advantages of HAPS in ITS applications is its ability to facilitate vehicle-to-infrastructure (V2I) communication. \cite{shinde2021towards} presents a new air-ground vehicle network architecture that integrates terrestrial and non-terrestrial edge computing platforms to address the limitations of traditional systems. It explores enabling technologies like federated learning, network softwarization, and computation offloading to support latency-critical and data-intensive vehicular services. Real-time transmission of data performed possibly by V2I systems allows vehicles to engage with road infrastructure, traffic management systems, and other connected vehicles, therefore improving decision-making and accident avoidance.
\cite{ren2021high} introduces a novel three-tier computation architecture for ITS by incorporating HAPS as stratospheric edge computing nodes. It uniquely combines HAPS-based data storage with intelligent task offloading and edge caching to address latency and scalability issues in autonomous transportation systems. 
Also, \cite{9685074} proposes a three-layered computational model using HAPS to improve ITS applications. The suggested model combines cars, RSUs, and HAPS to allocate tasks and cache effectively. Offloading computing tasks to HAPS helps the system lower latency in ITS services and increase processing efficiency. 
\cite{ren2023handoff} suggests a handoff-aware distributed computing framework for vehicular networks by integrating HAPS with RSUs and vehicles. The approach dynamically splits computing tasks among the three nodes to minimize delay and avoid interruptions caused by frequent edge network handoffs. An optimization model is developed to allocate bandwidth, power, and computing resources effectively and solved using a successive convex approximation method.
\cite{traspadini2022uav} also offers a queuing-theoretic model to optimize task offloading from ground vehicles to aerial platforms, minimizing processing time under various constraints. By modeling computation tasks as Poisson processes and applying queueing theory, it identifies the optimal offloading strategies that minimize latency while accounting for computational capacity and transmission delay.
Additionally, \cite{kharchenko2023study} focuses on the effects of parameters like transaction size, bit error rate (BER), and vehicle density on an HAP-RSU-vehicle architecture.  The evaluation is performed considering network load, packet error rate, and message delivery time.

\begin{table*}[!htbp]
  \scriptsize
  \renewcommand{\arraystretch}{0.5}
  \caption{Comparison of Advanced Aerial Communication \& Networking studies.
           HT = heterogeneous multi‑tier design;
           MB = explicit mobility handling;
           TECH = principal communication technology;
           FRQ = main operating band;
           EE = includes an energy‑efficiency metric or optimization
           }
  \label{tab:HAP-ITS}
  \centering
  \resizebox{\textwidth}{!}{%
  \begin{tabular}{|l|c|p{7cm}|c|c|c|c|c|}
  \hline
  \textbf{Study} & \textbf{Year} & \textbf{Contribution} & \textbf{Technology} & \textbf{Band} & \textbf{HT} & \textbf{MB} & \textbf{EE}\\
  \hline
    \cite{albagorynovel}            & 2020 & Applications of HAP for traffic monitoring and management      & WSN       & –         & \cmark & \xmark & \xmark \\\hline
    \cite{shinde2021towards}        & 2021 & Air–ground intelligent platform; open research challenges      & C‑V2X     & mmWave    & \cmark & \cmark & \xmark \\\hline
    \cite{ren2021high}              & 2021 & Three-layer joint offload–cache optimization                   & MEC       & 2 GHz     & \cmark & \xmark & \xmark \\\hline
    \cite{jaafar2022haps}           & 2022 & Three-tier HAPS-ITS architecture; Trans-Sahara case study      & MEC       & –         & \cmark & \xmark & \cmark \\\hline
    \cite{popoola2022exploiting}    & 2022 & Rural coverage \& capacity analysis with large HAP array       & mmWave    & 28/38 GHz & \xmark & \cmark & \xmark \\\hline
    \cite{traspadini2022uav}        & 2022 & Offloading factor for UAV/HAP-assisted VEC via queuing-theory  & mmWave    & 38 GHz    & \cmark & \xmark & \xmark \\\hline
    \cite{ren2023handoff}           & 2023 & Handoff‑aware task‑split (vehicle–RSU–HAPS) to cut delay       & 5G Edge   & –         & \cmark & \cmark & \cmark \\\hline
    \cite{shinde2023joint}          & 2023 & Joint air–ground FL cuts uplink overhead for CAV sensing       & 6G        & –         & \cmark & \cmark & \cmark \\\hline
    \cite{kharchenko2023study}      & 2023 & Traffic-load \& PER model for BS–HAP–RSU–Vehicle chain         & 5G   & –         & \cmark & \xmark & \xmark \\\hline
    \cite{djihane2023intelligent}   & 2023 & Comparative analysis of HAPS vs LEO latency and coverage       & FSO       & 1550 nm   & \xmark & \xmark & \xmark \\\hline 
    \cite{dutta2024fair}            & 2024 & Fairness‑oriented power/decoding optimisation                  & NTN       & S         & \cmark & \cmark & \cmark \\\hline
    \cite{naseh2024multi}           & 2024 & Federated‑split‑transfer learning across RSU–HAPS tiers        & 6G        & –         & \cmark & \xmark & \xmark \\\hline
    \cite{rzig2024dependency}       & 2024 & Dependency‑aware offloading in UAV–HAPS vehicular edge         & NOMA+MEC  & –         & \cmark & \cmark & \cmark \\\hline

    \cite{lian2021non}              & 2021 & 3-D wideband channel model for RIS-assisted HAP-MIMO links     & RIS       & 2 GHz     & \xmark & \xmark & \xmark \\\hline
    \cite{ji2024active}             & 2024 & Active RIS-enhanced NOMA scheme for HAP-MISO systems           & RIS+NOMA  & 2.4 GHz   & \xmark & \xmark & \xmark \\\hline
    \cite{ni2023beamforming}        & 2023 & RIS-aided beamforming \& interference cancellation for HAP-D2D & RIS       & –         & \xmark & \xmark & \xmark \\\hline
    \cite{alfattani2022beyond}      & 2022 & Beyond-cell coverage via HAPS-RIS architecture                 & RIS       & 2, 30 GHz & \xmark & \xmark & \xmark \\\hline
    \cite{9500697}                  & 2021 & GAT-based channel estimation for RIS-assisted HAPS backhaul    & 6G/RIS       & –         & \xmark & \xmark & \xmark \\\hline
    \cite{wu2024deep}               & 2024 & EE optimisation for RIS-aided satellite–aerial–terrestrial relay & RIS+FSO & 1550 nm / 2 GHz & \cmark & \xmark & \cmark \\\hline
    \cite{vegni2024enhancement}     & 2024 & RIS-enabled handover in 3-D ground–aerial–space networks       & 6G+RIS    & 1200 MHz  & \cmark & \cmark & \xmark \\\hline
    \cite{tanash2024integrating}    & 2024 & RIS for signal coverage and adaptability in HAP networks       & RIS       & –         & \xmark & \xmark & \xmark \\\hline
    \cite{an2024exploiting}         & 2024 & Multi-layer refracting RIS receiver for SWIPT-enabled HAP links& RIS+SWIPT & –         & \xmark & \xmark & \cmark \\\hline
    \cite{le2024harvested}          & 2024 & Energy-harvesting for FSO RIS-assisted ground–HAP–UAV system   & RIS+FSO   & 1550 nm   & \cmark & \xmark & \cmark \\\hline
    \cite{guo2023joint}             & 2023 & Joint optimisation for RIS-aided hybrid FSO SAGINs             & RIS+FSO   & 2.4 GHz   & \cmark & \xmark & \xmark \\\hline
    \cite{odeyemi2022reconfigurable}& 2022 & RIS-assisted HAPS relaying for multi-user links                & RIS       & –         & \xmark & \xmark & \xmark \\\hline
    \cite{alfattani2023resource}    & 2023 & Resource-efficient beyond-cell communications with HAPS-RIS    & RIS       & –         & \xmark & \xmark & \xmark \\\hline
    \cite{azizi2023ris}             & 2023 & Aerodynamic HAPS-RIS with multi-objective optimisation         & RIS       & 2 GHz     & \xmark & \xmark & \cmark \\\hline
    \cite{wu2023joint}              & 2023 & Beamforming for RIS-assisted satellite–HAP–terrestrial networks& RIS+FSO   & 2 GHz     & \cmark & \xmark & \xmark \\\hline

    \cite{aziz2014disaster}         & 2014 & Assesses effectiveness of LTE Release 8 HAPS for disasters     & LTE       & 1800 MHz  & \xmark & \xmark & \xmark \\\hline
    \cite{dong2015energy}           & 2015 & Energy-minimizing techniques via integrated HAP–satellite WSN  & WSN       & Ka/Ku,L/S & \cmark & \xmark & \cmark \\\hline
    \cite{almalki2017propagation}   & 2017 & Propagation channel model for aerial platforms in emergencies  & WiMAX/LTE &2.5/3.5 GHz& \xmark & \xmark & \xmark \\\hline
    \cite{Almalki2020}              & 2020 & ML-based rapid restoration of comms links via aerial platforms & WiMAX     & 2.5 GHz   & \xmark & \xmark & \xmark \\\hline
    \cite{qiu2021multi}             & 2021 & Layered clustering for multi-tier NTN disaster communications  & NTN       & –         & \cmark & \xmark & \xmark \\\hline
    \cite{baraniello2021application}& 2021 &  Rapid disaster response system    & FSO       & –         & \cmark & \cmark & \xmark \\\hline
    \cite{raj2024stochastic}        & 2024 & Markov model for energy efficiency in HAP–LEO systems          & NTN       & –         & \cmark & \xmark & \cmark \\\hline
    \cite{zhao2023backhaul}         & 2023 & Backhaul-constrained coverage analysis of HAP-LAP system       & mmWave    & –         & \cmark & \xmark & \xmark \\\hline
    \cite{andreadis2023role}        & 2023 & UAVs and HAPS for IoT-based monitoring in emergency scenarios  & LoRa+5G   &2 GHz/868 MHz& \cmark & \xmark & \xmark \\\hline
    \cite{baraniello2021application}& 2021 & The Application of HAPS for rapid disaster response            & LEO       & –         & \xmark & \xmark & \xmark \\\hline
    \cite{matracia2024unleashing}   & 2024 & Aerial RIS benefits in post-disaster scenarios                 & RIS       & –         & \xmark & \xmark & \xmark \\\hline
    \cite{yu2024research}           & 2024 & Field trials of UAV-based HAPS in air-heaven network           & 6G        & –         & \xmark & \xmark & \xmark \\\hline
    \cite{Alqasir2024}              & 2024 & Energy-efficient NTN with HAPs and UAVs for disaster areas     & NTN       & –         & \cmark & \xmark & \cmark \\\hline

    \end{tabular}}
\end{table*}

Another critical aspect of HAPS-enabled ITS is its support for autonomous and connected vehicles (CAVs). Such vehicles rely on large amounts of real-time data to make informed navigation, obstacle avoidance and traffic coordination decisions.  The study~\cite{albagorynovel} explores innovative traffic management applications for HAPs. It presents real-time traffic light control, accident detection, and vehicle tracking.~\cite{albagorynovel} focuses on vehicle tracking using HAP-based surveillance. HAPS enables edge computing functionalities that bring computational intelligence closer to vehicles, reducing the need for extensive data transmission to centralized cloud servers.~\cite{10597130} suggests GFSTL as a privacy-preserving and efficient learning framework for ITS. The study shows how HAPs can enhance data processing and decision-making for connected vehicles. HAPS can analyze huge transportation data and maximize vehicle movements using AI-driven analysis, thus improving safety and efficiency. Furthermore,~\cite{shinde2023joint} investigates the use of Federated Learning (FL) in ITS by integrating air-ground computing resources. It proposes a distributed FL framework that combines terrestrial networks, UAVs, and HAPS to optimize vehicular data processing while reducing communication costs.
Moreover,~\cite{naseh2024multi} introduces a novel distributed learning framework called Generalized Federated Split Transfer Learning (GFSTL) for 6G-enabled ITS. The approach combines Federated Learning, Split Learning, and Transfer Learning to address latency, privacy, and resource constraints in Aerial-Ground Integrated Networks (AGIN) using RSUs and HAPs. Simulation results show that GFSTL improves accuracy, reduces latency, and performs better than traditional learning methods, even in scenarios with limited or heterogeneous data.

Beyond urban settings, HAPa are significant for enabling ITS services in remote and rural areas. Implementing real-time traffic monitoring and emergency response systems on major highways is difficult. Because most of them do not have continuous network connectivity. \cite{jaafar2022haps} explores integrating HAPS into ITS frameworks for transcontinental highways. It underlines how limited conventional ITS solutions are in remote areas. It introduces real-time traffic monitoring, accident reporting, and vehicle tracking using HAPs. A case study of the Trans-Saharan highway shows the valuable advantages of HAPS in offering flawless connectivity and security improvements. Using HAPS-based networks, authorities can create seamless connectivity, improve road safety and minimize response times to accidents or breakdowns. 
\cite{djihane2023intelligent} presents a comprehensive analysis of how HAPs can enhance the performance of ITS in underserved or remote regions. The study compares HAPS with LEO satellites. It emphasizes HAPS's advantages in latency, coverage, and real-time connectivity. A case study of Algeria's 1020 km Highway of the Hauts Plateaux demonstrates the feasibility and impact of deploying tethered HAPS to improve traffic management, safety, and communication infrastructure.
\cite{cai2018location} proposes a resilience-focused clustering framework which  is tailored for HAP-assisted VANETs operating in dynamic and infrastructure-deficient environments. The LCCT-CA algorithm uniquely combines spatial location data and real-time mobility metrics. The study uses connection duration and neighborhood density to compute node stability for robust cluster formation. The algorithm significantly outperforms traditional schemes in emergency scenarios. Also, the algorithm offers superior reclustering speed and communication continuity. The integration of aerostats, UAVs and ground nodes into a unified hierarchical network sets a new direction for scalable, reliable and infrastructure-independent vehicular communication systems.

HAPs are also explored for next-generation aerial delivery systems. Enterprises interest in launching autonomous UAV-based delivery systems that require robust processing and communication infrastructure to monitor air traffic and coordinate flight paths. HAPs provide a consistent backbone for such uses, guaranteeing flawless communication between ground stations, UAVs, and logistical hubs. The study \cite{10794340} V2X communication architecture is designed to meet the stringent hyper-reliable and low-latency communication (HRLLC) demands of 6G networks. It addresses key challenges in autonomous driving, such as latency, connectivity, and energy efficiency, by proposing AI-enhanced mechanisms for traffic classification, data offloading, and system resilience. \cite{10794340} distinguishes itself by detailing a dual-network strategy and emphasizing green communication through solar-powered aerial platforms.

Regarding the environment, ITS run on HAPS can support projects aimed at sustainable mobility. ITS solutions can cut fuel use and greenhouse gas emissions by maximizing traffic flow and reducing congestion. Moreover, HAPS runs on solar energy, which makes it a green substitute for ground-based communication systems depending on a fossil-fuel-powered electricity grid.

Although including HAPS in ITS has several benefits, deployment, energy economy, and regulatory frameworks still present problems. \cite{dutta2024fair} optimization model that balances power distribution, interference management, and decoding order for improved network performance. The primary focus of current research is optimizing the lifetime of HAPS platforms, enhancing their energy storage capacity, and guaranteeing seamless handoff between several platforms to provide continuous service. Standardizing communication protocols also helps to enable HAPS to communicate easily with terrestrial and satellite networks.

\begin{figure*}
    \centering
    \includegraphics[width=0.97\linewidth]{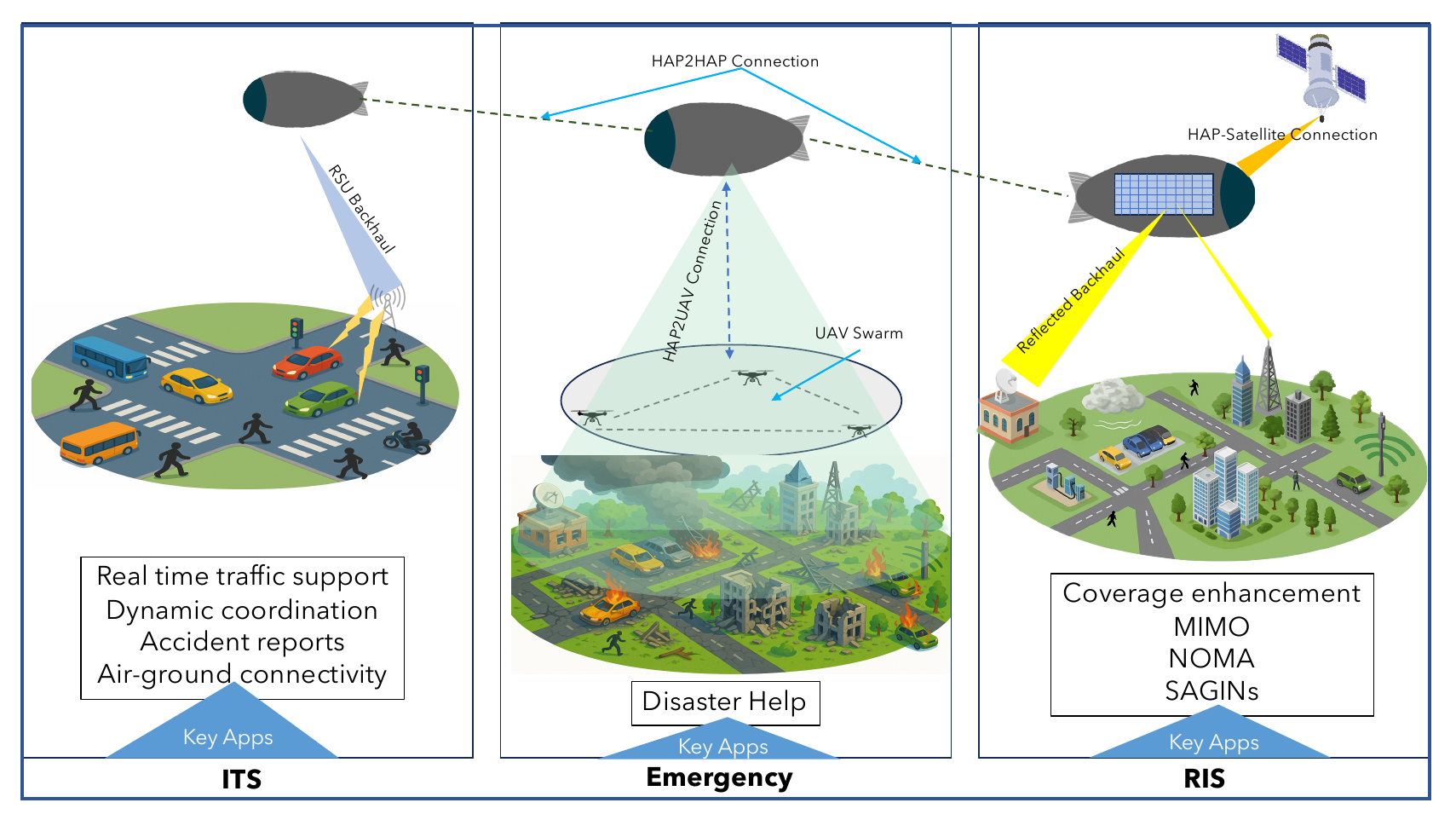}
    \caption{Advanced Aerial Communication and Networking Applications}
    \label{fig:enter-label}
\end{figure*}

\subsubsection{Disaster and Emergency Applications}
While foundational to modern connectivity, terrestrial communication infrastructures are often the first to fail in the wake of natural or artificial disasters; earthquakes, hurricanes, floods, wildfires, and armed conflicts can inflict severe physical damage on ground-based base stations, fiber-optic backbones, and power supplies. Under such conditions, partial infrastructure failure and unexpected traffic surges can paralyze communication networks and leave impacted populations without access to family contacts, medical support, emergency alerts, or rescue coordination channels. The collapse of communication hinders the immediate response of emergency services moreover, it obstructs longer-term recovery efforts and supply chain logistics.
HAPs have thus gained increasing attention as vital components in disaster-resilient communication ecosystems. HAPs serve as aerial communication hubs that can be rapidly deployed over disaster-stricken regions. Unlike satellites requiring substantial investment, lead time, and orbital synchronization, HAPs can be launched or redirected with far greater agility and at significantly lower cost.Their raised but non-orbital posture offers a steady line-of-sight (LoS) communication channel for ground users and other aerial or satellite nodes, enabling smooth integration into hybrid emergency networks.
The ability of HAPs to maintain wide-area coverage and relay real-time data from the ground to coordination centers makes them indispensable for enabling situational awareness in crisis environments~\cite{gharbi2019overview}. Equipped with advanced payloads, such as high-resolution cameras, thermal imaging sensors, and low-latency communication modules, HAPs can facilitate live mapping of disaster zones, track population movements, monitor environmental hazards, and guide UAVs or ground responders to high-priority areas. Moreover, they can re-establish backhaul links between isolated emergency service units and national response systems, even when traditional core networks are completely disabled~\cite{gharbi2019overview}.
Some recent deployments and pilot programs have demonstrated the feasibility of leveraging HAPs such as flying LTE or 5G base stations, dynamically allocating bandwidth, and prioritizing emergency communication channels. These platforms can also support interoperability between civilian and military units, or among international aid agencies operating in the same disaster area. Operating either independently or in concert with satellites and drones, HAPs provide a layered, robust, scalable communication architecture that guarantees continuity of operations even under highly demanding circumstances. Integrating HAPs into national and regional disaster preparedness systems is fast becoming a strategic need rather than a technological luxury as climate change raises the frequency and intensity of events occurring worldwide. Emphasizing their possible benefits and exceptional capacity in disaster and crisis communication, an overview study~\cite{gharbi2019overview} presents a thorough analysis of HAP networks. It addresses many facets including operational benefits, network architecture, challenges, and practical uses. It also provides detailed case studies that illustrate the effective use of HAPs in managing communication during crises.

HAPS's capacity to operate as airborne base stations is one of the main benefits of disaster response. Ground-based emergency teams and mobile devices can be connected to guarantee that voice and data connections remain functional. Mostly dependent on these skills are search and rescue activities, where real-time communication might mean the difference between life and death. Constant connectivity is vital for emergency responders to coordinate activities, properly allocate resources, and distribute critical information on affected areas.  Studies such as~\cite{aziz2014disaster,zhao2023backhaul} provide emergency networks and communication for disaster areas.                  
\cite{aziz2014disaster} assesses the effectiveness of LTE release 8 networks deployed through HAPS in disaster scenarios. Most dependent on these skills are search and rescue activities, where real-time communication might mean the difference between life and death. Constant connectivity is vital for emergency responders to coordinate activities, properly allocate resources, and distribute critical information on affected areas. The paper \cite{zhao2023backhaul} extensively analyzes the coverage capabilities of aerial communication systems integrating both HAP and LAP in post-disaster environments. It focuses explicitly on how backhaul constraints limit coverage effectiveness and presents optimization methodologies to overcome these challenges. Cooperating with HAPs \cite{matracia2024unleashing}, investigates the benefits of aerial RIS (ARIS) technology in post-disaster communication by comparing it with terrestrial RIS (TRIS) and aerial base stations (ABS). The study demonstrates that ARIS significantly enhances coverage and communication quality by supporting surviving terrestrial base stations. ARIS is shown to play a critical role in rapidly restoring communication infrastructure after disasters.

HAPs are important for disaster assessment and response planning due to their imaging and sensing capabilities. HAPs can use synthetic aperture radar (SAR), thermal imaging sensors, and high-resolution cameras to capture images of disaster areas. This allows for the detection of damage caused by the disaster and the location of survivors. Unlike satellite-based monitoring systems, which might have restricted revisit intervals, HAPs can constantly observe impacted areas, giving decision-makers ongoing situational knowledge. For monitoring, paper~\cite{andreadis2023role} examines the use of aerial platforms such as UAVs and HAPS to support IoT applications in emergency management scenarios. It outlines how IoT data collected by these aerial platforms can be efficiently transmitted to Emergency Management Centers (EMC) to facilitate rapid responses. The advantages and limitations of NB-IoT and LoRa technologies are compared, especially in the context of their use in reliable communication protocols during emergencies. Furthermore, \cite{baraniello2021application} proposes deploying HAPS to obtain rapid, high-resolution imagery immediately following disasters. It highlights HAPS' advantage in providing persistent surveillance due to their ability to maintain station-keeping positions at altitudes of 18-20 km, offering significantly better spatial resolution and time persistence than traditional satellite imagery.

In places devastated by disasters where traditional communication systems have been disrupted, HAPs can also help to enable online connectivity. For affected populations, they can provide text messaging, internet connectivity, and emergency calls acting as makeshift cell towers. First responders gain from this capacity, but so do people who can call loved ones, get emergency information, and ask for help as needed. Rebuilding communication infrastructure using HAPs will help to boost disaster resilience and support for recovery initiatives significantly. \cite{liu2019energy} develops a new propagation channel model specifically tailored to aerial platforms used during emergencies. It aims to optimize ad hoc network performance and reliably predict received signal strength (RSS) for rescue operations. \cite{qiu2021multi} focuses on multi-tier aerial and space platforms structured into clustered networks to enhance disaster communication performance. It proposes a layered clustering method that improves network efficiency, scalability, and robustness in emergency scenarios. \cite{santovito2020ka} presents the design, development, and testing of specialized up/downconverter hardware operating in the Ka-band frequency for HAPS. The hardware aims to support robust and reliable emergency aid communications, providing technical solutions to challenges associated with frequency conversion and signal integrity. Detailed performance assessments and practical implications for emergency scenarios are discussed.
\cite{Almalki2020}, suggests a novel approach for rapidly restoring communication networks in disaster-stricken areas by utilizing aerial platforms. These platforms use an ad-hoc network model to optimize air-to-ground propagation models. The framework essentially improves connectivity, lowers power consumption, and increases network dependability during emergencies with using different criterias like Quality of Service (QoS) and Grade of Service (GoS).

HAPS's rapid adaptability and flexibility are other important advantages. Unlike ground-based infrastructure, which can take a long time to deploy, HAPs can be operational within hours. They can be repositioned when necessary. HAPs can be integrated with existing communication technologies to create a comprehensive emergency response system.
For example, \cite{almalki2017propagation} proposes new approaches to increase the energy efficiency of integrated HAP-satellite communication systems designed for emergency communication situations. The study focuses on transmission techniques that reduce energy consumption while maintaining communication quality. On the other hand, the study~\cite{raj2024stochastic} presents a stochastic model using the Markov Regenerative Process to analyze and improve energy efficiency in HAP and LEO integrated systems. The model includes operational states that save energy. It also focuses specifically on solar-powered HAPs, and evaluates the balance between energy harvesting and energy saving mechanisms. \cite{yu2024research} covers the potential use of UAV-based HAPS in 6G networks, thus highlighting the relevance of these systems in post-disaster conditions.

HAPs are  transformative solutions for protecting and restoring communication infrastructure after disasters. Their wide coverage area, rapid deployment, and flexible connectivity capabilities enable them to play a vital role in emergency response situations. HAPs can provide real-time coordination while increasing situational awareness. This enables communication for both first responders and affected communities. 

\subsubsection{Reconfigurable Intelligent Surfaces}

As wireless communication systems evolve toward 6G and beyond, the limitations of terrestrial infrastructure in providing ubiquitous, low-latency, and energy-efficient connectivity have prompted a paradigm shift toward NTNs~\cite{10250790}. High HAPs have emerged as a promising solution to bridge connectivity gaps in rural, remote, and disaster-stricken regions. At the same time, RIS has attracted a lot of interest for its capacity to programmable propagation by means of software-controlled reflections, so modifying wireless environments~\cite{tekbiyik2022reconfigurable}. HAPs and RIS technologies together offer a strong architecture capable of achieving adaptive coverage, improved spectral efficiency, and intelligent resource allocation. This synthesis of  RIS-assisted HAP systems is shaping the future of wireless communication through innovations in channel modeling, beamforming, hybrid networking, energy optimization, and network intelligence.

HAPs, are ideal for providing wide-area coverage in rural, remote, or emergency areas. However, they still suffer from signal degradation due to atmospheric disturbances, mobility dynamics, and LoS limitations. With its ability to passively or actively reflect and manipulate electromagnetic waves, RIS can reshape wireless propagation environments in real time.
Several studies in the literature propose RIS-enabled channel models to characterize HAP scenarios better. For instance, in \cite{lian2021non} a 3D non-stationary wideband channel model tailored explicitly for HAP-MIMO systems captures the spatial and temporal variability introduced by mobile nodes and multi-path environments. Incorporating the special scattering and reflection properties of RIS elements, the model faithfully records spatial and temporal fluctuations in the wireless channel resulting from user mobility and HAP platform movement. Strategically placed on HAPs, RIS panels dynamically control reflections to reduce blockage effects and maximize propagation conditions.
The study \cite{9500697} develops an advanced channel estimation approach utilizing Graph Attention Networks (GAT) in full-duplex RIS-assisted HAPS for wireless backhauling. RIS elements are strategically positioned on HAPS platforms to reflect signals, overcoming propagation challenges and enabling full-duplex operations. The proposed graph-based estimation technique effectively models the complex interdependencies between multiple RIS reflections and aerial paths, significantly enhancing accuracy and robustness in dynamic channel environments.
\cite{ji2024active} examines the performance gains achievable by integrating active RIS into HAP-based Multi-Input Single-Output (MISO) systems NOMA. Capable of signal amplification, active RIS elements are placed on HAPs to improve received signal strength and guarantee fair user resource distribution. Targeting phase-shift configurations and beamforming design to maximize feasible user rates under limited power resources, a joint optimization framework is presented.

RIS introduces intelligent reflectivity that significantly enhances beamforming precision and interference cancellation capabilities.  RIS enables directional signal focusing in HAP-based Device-to-Device (D2D) and Non-Orthogonal Multiple Access (NOMA) systems, reducing unwanted leakage and improving user fairness.
Several papers propose SLNR-based joint precoding, beamspace signal alignment, and interference mitigation schemes that demonstrate the efficacy of RIS in dense or mobility-heavy deployments.  The multi-objective optimization of RIS configurations offers substantial improvements in throughput and system reliability.
\cite{ni2023beamforming}, explores RIS integration in HAP-based D2D communications to address challenges associated with interference and beam misalignment.  RIS arrays are deployed on HAPs to dynamically shape propagation paths, mitigate inter-device interference, and facilitate direct high-capacity links between terrestrial devices.  The proposed joint optimization method greatly reduces interference and improves overall communication reliability by simultaneously selecting ideal RIS phase shifts and beamforming vectors. As a result, the proposed approach provides a good structure to efficiently manage interference and improve connectivity in RIS-enabled HAP-D2D communication networks.

Additionally, \cite{ji2024slnr} presents a new joint precoding strategy for RIS-supported beamspace HAP systems using NOMA. The authors formulate an SLNR-based precoding criterion to optimize signal strength for users. RIS units assist in dynamically reshaping spatial beams from HAPs.  This solution increases coverage in design conditions prone to overload or interference. In the interdisciplinary work \cite{azizi2023ris}, the authors propose a joint optimization approach for RIS-equipped HAPS. The solution investigates the intersection of communications and aerodynamics. The platform manages electromagnetic signal steering via RIS. Additionally, it adapts the flight path to optimize coverage area and signal transmission.

Several contributions emphasize the multi-layered architecture of modern non-terrestrial networks, often described as Satellite-Aerial-Ground Integrated Networks (SAGINs).  Here, RIS is used both at the HAPs and in ground stations, UAVs, and even in satellite relays to optimize Free-Space Optical (FSO) links and RF channels.

In hybrid FSO/RF systems, RIS mitigates the effects of atmospheric turbulence and pointing errors, ensuring stable links under harsh propagation conditions.  Deep reinforcement learning (DRL) algorithms are often employed to dynamically adapt RIS configurations based on real-time environmental inputs, traffic loads, and energy budgets.
\cite{wu2023joint} suggests  joint optimization strategy for RIS-enhanced FSO links in SAGINs, using DRL. Deployed on mobile HAPs, the RIS modules are trained via DRL to adaptively align beams and change reflection angles to counter turbulence and misalignment in the FSO channel.  The integration reduces power consumption, enables dynamic relay selection, and effective link switching.  By using intelligent learning agents, the proposed RIS-HAP framework enables real-time FSO management across heterogeneous 3D networks. 
Similar to \cite{wu2023joint}, \cite{guo2023joint} enhances SAGIN performance via RIS-equipped HAPs and DRL-powered optimization.  The RIS units compensate for FSO's high sensitivity to alignment and weather by dynamically adjusting their surfaces.  A deep learning model learns optimal policies for beam routing and node coordination, taking into account atmospheric variance. Compared to traditional static beam systems, RIS-HAP systems increase efficiency and stability at different altitudes. The approach is therefore applicable for multimedia and data-intensive applications from space to the ground. The study \cite{ye2022nonterrestrial} investigates how RIS can support communication between HAPs, UAVs, and ground stations. The system model incorporates RIS in both static towers and mobile aerial vehicles. This enables flexible real-time signal redirection.  


Energy efficiency is critical for long-duration, high-altitude missions in HAP-based systems. Many articles propose RIS-supported solutions that optimize Simultaneous Wireless Information and Power Transfer (SWIPT) using RIS. RIS structures enable selective beamforming to direct information to users or energy to harvesting modules. Researchers present DRL-based energy optimization frameworks. The frameworks also reduce power consumption without compromising communication quality. These findings demonstrate the importance of RIS for developing sustainable HAP networks.
\cite{wu2024deep} investigates RIS deployment in satellite, airborne, and terrestrial relay networks. RIS dynamically enhances signal propagation. HAPs are strategically employed to mitigate congestion effects between RIS layers. A DRL-based algorithm optimizes RIS configurations, relay placement, and power allocation to ensure energy consumption efficiency. \cite{wu2024deep} offers an alternative solution for energy-efficient communication infrastructure with airborne and satellite components.
In addition, \cite{sun2023scalable} tackles the challenges of signal and energy transfer in SWIPT systems by utilizing multi-layer refracting RIS on HAPs.  A scalable beamforming approach is developed to ensure system robustness under uncertain CSI and dynamic aerial conditions.  The RIS architecture enables precise directional control, steering RF waves toward different functional modules (information decoder vs. energy harvester).  The proposed scheme guarantees stable performance with respect to user density, mobility, and energy needs.  \cite{sun2023scalable} highlights RIS-HAP synergy as a cornerstone for future self-sustaining wireless energy delivery networks.
Other study \cite{an2024exploiting} proposes a unique RIS-assisted receiver architecture that utilizes multi-layer refracting surfaces to enable efficient SWIPT in HAP-supported systems.  The RIS layers are configured to selectively direct incoming RF energy toward information decoding or energy harvesting modules.  A joint optimization model is presented for RIS phase shifts and power allocation, balancing communication quality with harvested energy levels.
The paper \cite{le2024harvested} investigates a hybrid system where FSO links, aided by RIS, connect ground nodes to HAPs and UAVs for both data transmission and energy harvesting.  The RIS elements are optimally placed to redirect and focus FSO beams under atmospheric turbulence and fading, improving beam alignment and efficiency.  Analytical models are developed to evaluate the harvested energy under composite Gamma-Gamma and pointing error distributions.  The study compares direct FSO transmission with RIS-aided routing, showing that RIS substantially increases harvested energy and link robustness.
With a different perspective, \cite{alfattani2021link} presents a full-scale link budget analysis tailored for aerial platforms using RIS.  By incorporating RIS into HAP payloads, the study quantifies the gain in beamforming efficiency, path loss mitigation, and energy usage.  The authors analyze how varying RIS configurations impact link budget under different atmospheric and altitude conditions.   The analysis reveals that RIS allows fine-grained control over energy distribution and enables HAPs to serve larger areas without excessive power scaling.

Furthermore, one of the most exciting applications of HAP-RIS integration is transcellular communication. Traditional cellular networks are limited by fixed infrastructure and terrain constraints. By deploying RIS-equipped HAPs, signal beams can be intelligently redirected to fill coverage gaps, connect isolated communities or maintain connectivity continuity during natural disasters or infrastructure failure.
Studies show that RIS helps shape adaptive coverage zones, allowing a single HAP to replace multiple terrestrial base stations. 
\cite{alfattani2022beyond} focuses on extending communication coverage beyond traditional cellular boundaries through RIS-equipped HAPS.  By mounting RIS on HAPS, dynamic beam control is achieved, allowing signal reflection toward distant or obstructed areas without infrastructure.  A unique optimization algorithm dynamically adjusts RIS configurations based on user distribution and coverage requirements, maximizing signal strength and connectivity reliability.  Simulation outcomes showcase extensive coverage improvement and stable high-quality links in remote regions, underscoring RIS-enabled HAPS as a feasible solution for rural broadband access and emergency network deployments.
\cite{alfattani2023resource} addresses the problem of extending cellular coverage beyond traditional cell boundaries using HAPS platforms equipped with RIS. For rural or emergency installations, his proposed architecture is perfect since it emphasizes on low resource consumption while preserving high coverage quality.  The system can guide beams towards disconnected or low-signal areas by dynamically changing the phase shifts on RIS placed on HAPS.  An algorithm for optimization is developed to balance surface use efficiency, bandwidth, and energy consumption.
In addition tho these studies, \cite{tanash2024enhancing} explores the structural and operational enhancements of HAP-based networks by incorporating Reconfigurable Intelligent Surfaces.  RIS panels, attached to both airborne and auxiliary nodes, intelligently reshape wireless propagation to counteract path loss and shadowing effects in the stratospheric layer.  The system design supports adaptive reflection control, enabling the network to respond to environmental changes and user mobility.

Managing handover processes between aerial, terrestrial, and satellite nodes remains a key challenge in vertical heterogeneous networks.  RIS has shown promise in mitigating handover failures by intelligently redirecting signal paths during transitions.  Some studies integrate sensor data into RIS control loops to enable real-time environment-aware reflection, optimizing user experience and reducing latency.
Several papers also focus on DRL-driven autonomous RIS control in mobile settings.  Here, agents learn optimal reflection patterns, handover timings, and relay switching strategies, contributing to a fully intelligent, self-healing network fabric. 
\cite{vegni2024enhancement} tackles the issue of handover interruptions in multi-tier 3D networks by leveraging RIS deployed on HAPs.Handovers can cause latency and link degradation in complex settings where users negotiate ground, aerial, and even satellite layers.  RIS makes smooth node transitions by dynamically guiding signal paths and reconfiguring reflection parameters in real-time.  The proposed solution enables predictive handover decisions based on mobility patterns and network topology awareness.
For improving connectivity, the study \cite{tanash2024integrating} explores the seamless integration of RIS within HAP networks to address connectivity challenges in obstructed and dynamic environments.  The RIS panels are deployed on both HAPs and nearby reflective surfaces to intelligently redirect signals, particularly in urban canyons and remote rural zones.  The system offers adaptive link reinforcement for users suffering degradation resulting from mobility or terrain by means of real-time beam manipulation.  Optimizing models for RIS configuration are presented in this work to guarantee low-latency and strong communication links remain. 
One comparative study contrasts aerial IRS nodes mounted on HAPs with terrestrial IRS in terms of performance, coverage, and deployment complexity.  \cite{shaik2024performance} presents a comparative analysis of aerial vs. terrestrial IRS deployments in integrated satellite–HAPS–ground systems.  Aerial IRS, mounted on HAPS, provides line-of-sight flexibility and broader angular control compared to fixed ground IRS.  The study evaluates throughput, latency, and energy efficiency under dynamic topological scenarios.

\subsection{Integrated Sensing} \label{section2}
HAPs have become essential components of next-generation sensing and monitoring systems in the last few years. It is particularly valid in areas that are large, barely interconnected, or that are likely to change rapidly.   HAPs operate within the stratosphere and provide continuous aerial surveillance, wide-area coverage, and clear LoS communication links. This contributes to them very useful for many different fields, including large-scale environmental monitoring, precision agriculture, disaster response coordination, and maritime domain awareness. HAPs can be set up and moved around quickly, unlike satellites, which makes them more flexible and adaptable, and costs less to run and has less latency. Their high vantage point enables flawless interaction with ground-based low-power IoT devices, enabling continuous data collection across heterogeneous and sometimes isolated sensor networks. Their high vantage point enables flawless interaction with ground-based low-power IoT devices, enabling continuous data collection across heterogeneous and sometimes isolated sensor networks.

This synthesis builds on an increasing corpus of multidisciplinary research converging on the intelligent use of HAPs for distributed sensing. It emphasizes how IoT architectures with energy-efficient sensors can be integrated, as sensor nodes periodically interact with HAPs to save power and guarantee timely updates. Moreover, federated learning techniques allow decentralized model training across sensor networks without centralizing sensitive raw data, preserving bandwidth and privacy. Reinforcement learning algorithms are being applied to optimize sensing schedules, trajectory planning, and dynamic resource allocation, allowing HAPs to adapt to changing environmental or operational conditions in real-time. Furthermore, hierarchical processing frameworks are emerging. Raw data is preprocessed at edge layers before being forwarded to centralized platforms for deeper analysis and storage.

\begin{figure*}
    \centering
    \includegraphics[width=0.99\linewidth]{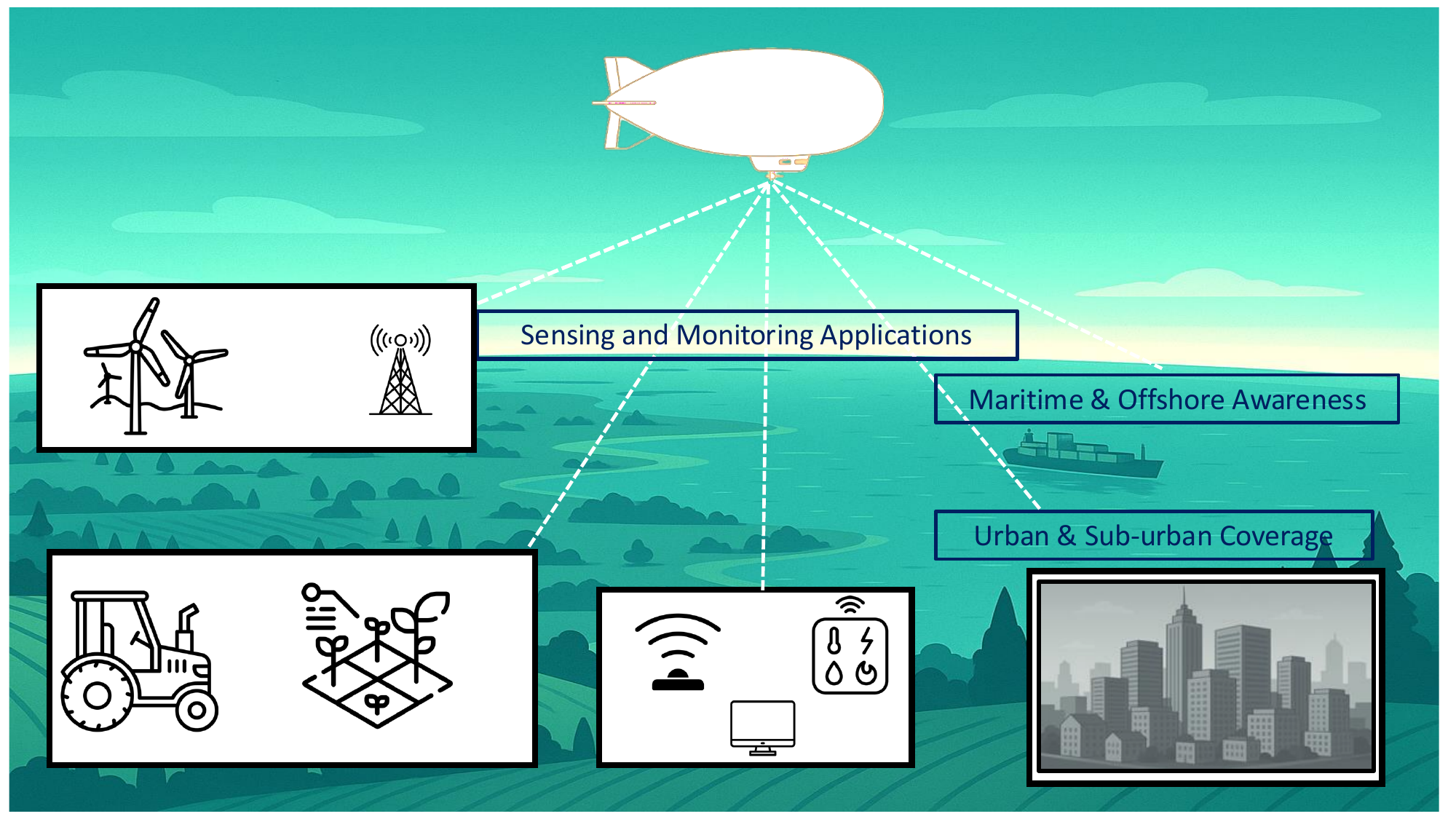}
    \caption{Integrated Sensing and HAPs}
    \label{fig:section3}
\end{figure*}

 Beyond single-mode observation, recent implementations explore multi-modal sensing capabilities where HAPs fuse data from visual, thermal, spectral, and atmospheric sensors to enable richer environmental intelligence. These systems improve situational awareness and support early warning mechanisms for natural disasters such as wildfires, floods, and oil spills. As a result, HAP-enabled sensing infrastructures are increasingly viewed as vital components of resilient innovative ecosystems. The combined understanding gained from the current literature supports the vital contribution of HAPs in changing the scalability, responsiveness, and intelligence of monitoring systems in several fields as shown in Figure \ref{fig:section3} .

\subsubsection{Sensing and Monitoring Applications with HAPs}
Numerous pioneering studies illustrate the transformative potential of HAPs in acting as airborne gateways or intelligent relay hubs for environmental sensing and agricultural resource management, particularly in remote, rural, or disaster-prone regions where terrestrial communication infrastructure is limited or entirely absent. Leveraging their quasi-stationary positions in the stratosphere, HAPs can maintain persistent connectivity with ground-level Wireless Sensor Networks (WSNs) or widely dispersed IoT-enabled devices that monitor vital environmental indicators such as soil moisture, temperature gradients, humidity levels, air quality, and crop health metrics. These aerial systems distill acquired data to edge or cloud servers for real-time analytics, decision-making, and predictive modeling, so closing the gap between local sensing and centralized data processing. Moreover, their extensive line-of-sight coverage helps to coordinate diverse sensor clusters over large distances, so supporting scalable and reasonably priced monitoring systems. In addition to improving situational awareness and early anomaly detection, including HAPs in agri-environmental systems helps to enable precision agriculture methods that maximize irrigation, fertilization, and pest control strategies, contributing to higher crop yield, resource efficiency, and climate-resilient farming.

LoRa and other low-power communication technologies can be effectively combined with HAP-based architectures to support smart agriculture and environmental sensing. By acting as mobile aerial gateways, HAPs can collect data from distributed sensor nodes across large rural areas and relay it to cloud platforms for processing and visualization. This approach significantly reduces the dependency on dense ground infrastructure while enhancing coverage and reliability in remote or hard-to-reach regions. 
\cite{9911442} investigates the feasibility of using LoRa technology for IoT applications in remote areas by utilizing HAPS. It analyzes the trade-offs between HAPS altitude, IoT device transmission power, and device density, considering factors like shadowing and collisions to validate the applicability of LoRa for HAPS communications. 
\cite{almarhabi2021lora} suggests using LoRa systems mounted on HAPS)to cover larger areas and presents a link budget analysis for LoRa operating at 868 MHz, considering path loss based on elevation angle in different urban scenarios. The study optimizes the HAP altitude to maximize radio coverage for each urban environment with varying LoRa spreading factors.
\cite{uyanik2022investigation} investigates the performance of LoRaWAN-based communication between ground-level IoT devices and HAPS, which serve as mobile aerial gateways. Using simulations built on the ns-3 platform and real-world hardware specifications, the authors explore how key parameters such as the number of devices, their distribution area, the speed of HAPS, and LoRaWAN duty cycle settings impact the success and failure rates of data transmission. The study also introduces regression-based equations to predict communication performance, helping designers make informed decisions before real-world deployment.  
\cite{andreadis2022lowpower} introduces an opportunistic protocol for UAV-assisted IoT data collection in remote areas, synchronizing sensor transmissions with UAV passes to optimize energy efficiency. A preliminary testbed validates LoRa propagation conditions for ground sensors.

Furthermore, intelligent irrigation systems can benefit from HAP-supported architectures where sensor data is transmitted to aerial platforms for centralized analysis. HAPs can compute and deliver optimized irrigation schedules back to the ground network, enabling autonomous and efficient water resource management based on environmental inputs such as soil moisture, temperature, and humidity. Furthermore, integrating web-based monitoring platforms with HAP-based WSNs provides farmers and city planners real-time insights into pollution levels, weather conditions, and irrigation efficiency. 
\cite{kusuma2019design} designs and implements a HAPS-based air pollution monitoring system using Wireless Sensor Networks (WSN). HAPS enhances WSN capabilities by acting as a central collection and data processing point for sensor nodes. The system uses a Raspberry Pi 3 as the control center and a microstrip patch antenna for wireless communication.
Similar to \cite{kusuma2019design}, \cite{adha2019design} presents a real-time air pollution monitoring system that utilizes WSNs integrated with High HAPs and a web-based application interface. The system gathers environmental data such as CO and PM10 levels through distributed sensors, which transmit their readings to a HAPs-based server. TThe study demonstrates the system's ability to operate in real time, providing a scalable and user-friendly platform for air quality monitoring in urban areas. This approach supports ecological management and policy-making by increasing public awareness and accessibility to environmental data.
In \cite{albagory2018novel} the authors propose a novel irrigation WSN using a HAP and an adaptive switched-beam concentric circular antenna array to provide wide area coverage and improve communication performance for irrigation control. The system aims to bridge the network deployment gap in remote areas lacking communication infrastructure. In these scenarios, the energy efficiency and communication reliability of IoT devices become critical. HAPs alleviate these concerns by offering better LoS and lower transmission power for sensor nodes with extending their battery life and reducing communication latency.

Another promising application area is real-time surveillance for observation. HAPs can rapidly deploy over affected zones or critical water routes, providing high-resolution images, thermal readings, and atmospheric data even when terrestrial or satellite networks are compromised.
In maritime domains, optical sensing from HAPs can monitor ship routes, detect oil spills, or observe illegal fishing activity. 
\cite{brauchle2021towards}  presents a design for an integrated optical payload system tailored for HAPs to provide new applications in maritime remote sensing safety and security. The proposed system includes a gimbal, refractive optics, a 150 MPix sensor, a computational stack for AI-based object detection, and two data links for control and transmission. The design aims to meet the challenging weight, power, and data processing requirements of HAP-based maritime surveillance. 
For sensing devices, \cite{li2023high}  proposes a hierarchical computing offloading framework for marine-IoT networks, utilizing HAPs and a hybrid NOMA-FDMA transmission scheme to minimize overall delay. The framework involves offshore sensing devices offloading computation to HAPs via NOMA and HAPs further offloading to an onshore base station via FDMA, with an SCA-based algorithm designed to optimize the process. 
\cite{Rees2020} describes the development of a lightweight, high-resolution surveillance camera designed for deployment on HAPs. The camera is designed to operate at ~20 km altitude and achieve a ground resolution of better than 120 mm, utilizing a range of materials to minimize weight and match thermal expansion.
Disaster areas also benefit from multi-modal sensor integration, where HAPs gather data from mobile ground units (e.g., drones or vehicles), providing a unified aerial perspective. In such scenarios, the combination of thermal, optical, and RF sensing enables comprehensive situational awareness, enhancing the efficiency of rescue missions and environmental mitigation efforts.
Furthermore, \cite{zheng2023analysis} examines the positioning performance of a HAPS-aided GPS system in urban areas using both simulation and physical experiments, demonstrating that HAPs can improve positioning accuracy and the applicability of a RAIM algorithm. The study models HAPS measurements and shows improvements in HDOP, VDOP, and 3D positioning accuracy, especially in dense urban environments.

More recent studies explore incorporating machine learning algorithms in HAP-assisted sensing systems to optimize data collection strategies. To enhance privacy and data security, federated learning can be employed in HAP-assisted sensing networks. Edge devices can train models locally and share only recent updates instead of aggregating raw sensor data centrally, so enabling the HAP to support cooperative learning while maintaining data confidentiality. Operating as a FL aggregator, the HAP safely gathers model updates from edge devices while preserving user privacy. \cite{wang2024covert} proposes a covert communication-enabled air-ground integrated federated learning (AGIFL) network for urban sensing. It introduces a friendly jammer UAV to enhance information security and minimizes federated learning latency by jointly optimizing user transmit power, jammer power, and local training accuracy.

DRL techniques can be applied in multi-layered network environments involving HAPs to optimize data freshness and collection frequency. Such learning-based strategies help prioritize critical updates, especially in time-sensitive sensing scenarios, by dynamically adjusting HAP behaviors based on network feedback. Here HAPs act as an intermediary, gathering time-sensitive data from IoT devices and forwarding it to satellites or ground stations. By means of constant learning, reinforcement agents embedded in HAP systems can independently identify optimal data collecting and transmit strategies, so guaranteeing the timely delivery of high-priority or contextually relevant sensor data. \cite{Zhang2025AoI} introduces a SAGIN-based IoT system where UAV trajectory design and network configuration are jointly optimized using deep reinforcement learning and a matching game to minimize the age of information (AoI) under cost and practical constraints.

HAPs have great potential for sensing uses; hence, several system-level issues have to be resolved to guarantee their efficiency. One main problem is the limited power and payload capacity; although HAPs can stay airborne for long times, their weight limits them and calls for careful scheduling and the use of lightweight sensors. Another challenge is the trade-off between latency and bandwidth especially when sending high-resolution data like video or hyperspectral images, which can overwhelm HAP-to--ground communication links. Edge computing at the HAP level is proposed in hybrid architectures meant to support this. Resilience and fault tolerance are also rather crucial, particularly in dynamic weather or disaster scenarios when HAPs must remain constant and adapt to surroundings. At last, security and privacy concerns are quite important, especially in urban or tactical surveillance systems; hence, preserving data integrity and user anonymity demands for efficient solutions including federated learning and strong encryption techniques.

\subsubsection{Reliable Communication and IoT Infrastructure with HAPs}
Designing and implementing communication infrastructures that are scalable, resilient, and inclusive of rural, remote, and geographically isolated communities is critically important as the need for ubiquitous, high-quality connectivity continues to expand outside of urban areas. The high cost, logistical difficulties, and environmental restrictions of deploying terrestrial infrastructure frequently pose serious problems for these underserved areas. HAPs have become a game-changing technical solution able to close the digital divide by offering high-capacity, low-latency communication services to places traditional networks cannot reach.
HAPs serve as intelligent relay nodes or quasi-stationary aerial base stations operating at roughly 18 to 25 kilometers in the stratosphere, providing reliable LoS connectivity and wide-area wireless coverage. HAPs are particularly useful for on-demand coverage in emergency and long-term infrastructure scenarios because they can be deployed quickly, repositioned dynamically, and maintained at a fraction of the cost of satellites. Future multi-tiered network architectures that integrate terrestrial, aerial, and space-based assets will require them to easily communicate with ground-based IoT devices, mobile users, and terrestrial base stations due to their high yet reachable altitude.
The feasibility of HAPs as independent and self-sustaining network nodes has been further improved by recent developments in solar-powered endurance systems, lightweight payload technologies, and AI-driven network management. HAPs enable real-time data exchange, local content caching, and long-distance backhaul connectivity in rural and isolated areas. This allows for basic internet access and more sophisticated services like environmental monitoring,  precision agriculture, and remote education. Furthermore, integrating with low-power IoT networks enables large-scale, sustainable sensor deployments without needing dense terrestrial gateways.
Future 6G and non-terrestrial network (NTN) paradigms, where connectivity is anticipated to be continuous, context-aware, and widely available, are also made possible by HAPs' support for both vertical (ground-to-air) and horizontal (air-to-air, air-to-ground) communication layers. By means of their purposeful placement over digital deserts, empowerment of impoverished communities, global advancement of digital equity, and fastening of socioeconomic development, they can significantly change maps of global connectivity.

One of the primary benefits of HAP-assisted communication infrastructure is its ability to extend coverage to regions where terrestrial networks are either sparse, economically unfeasible, or damaged (e.g., during natural disasters). HAPs can offer persistent connectivity to ground-based IoT devices, forming aerial links that enable data exchange with cloud services or other network tiers. This architecture is particularly beneficial in large-scale IoT deployments such as precision agriculture, environmental monitoring, or rural healthcare services. 
\cite{10318836} investigates the use of High HAPs to provide communication services in rural areas, focusing on the efficient allocation of limited resources between freshness-aware real-time services (measured by AoI) and conventional services (measured by data rate). The study investigates the possibility of FSO backhaul to improve performance and suggests both static and dynamic resource allocation schemes, with DRL being used for the latter.
However, as an alternative standalone network to offer seamless connectivity and coverage in remote areas, \cite{sibiya2019reliable} suggests a dependable IoT network architecture. Potential uses of the suggested architecture are illustrated in a number of industries, including smart cities, healthcare, logistics, and agriculture.

Through wise resource allocation, HAPs dramatically increase reach and improve communication performance. Beamforming techniques, for example, can assist HAPs in directing their transmission towards particular clusters of IoT nodes or users. These techniques may be improved by AI models like LSTM or deep learning. This lowers power consumption and interference while increasing signal strength and spectral efficiency. \cite{xiao2019lstm} suggests a novel IoT network utilizing HAPs and addresses the challenge of maintaining reliable connections between ground gateways and user equipment due to HAPs' susceptibility to various factors. To overcome this, the paper proposes a Direction of Arrival (DoA) prediction method based on the LSTM model. \cite{hu2024kaband} uses LSTM to predict the location of UAVs to improve beamforming in 5G IoT networks. The study presents a system plan for a Ka-band up-down converter for emergency aid in a HAPS. The proposed system aims to replace damaged base stations by up-converting cellular signals to the Ka-band for a direct link to a ground station and down-converting back to cellular signals.

To effectively control the data flood from many sensors and smart devices, researchers have investigated hierarchical architectures and clustering methods. HAPs in these designs receive data from ordered clusters of ground devices either as sink nodes or intermediate aggregators.
\cite{wang2022performance} evaluates the performance of different Low-Power Wide Area Network (LPWAN) technologies (LoRa, NB-IoT, SigFox) in NTN scenarios involving UAVs, HAPs, and LEO satellites. It also formulates an optimization problem to determine whether IoT sensors should offload traffic to LEO satellites to reduce congestion on terrestrial gateways.
\cite{mahyastuty2020clustering} proposes a clustering algorithm for WSNs using HAPs in 5G networks, addressing the challenge of HAP movement. The algorithm adaptively re-clusters by selecting new cluster heads from sensor nodes within the updated HAP coverage area, minimizing the number of unconnected sensor nodes.
In addition, \cite{gharib2024high} implements a HAPS-greedy clustering (HAPs-GC) scheme for WSNs to support massive IoT applications, considering both connectivity between sensor nodes and their connectivity with HAPS. Simulation results demonstrate that HAPs-GC significantly increases WSN throughput while maintaining similar energy consumption compared to existing HAPs-based schemes. 
\cite{pangestu2021parameter} offers a parameter-based clustering algorithm for WSNs using High HAPs as base stations, aiming to improve energy efficiency by considering remaining energy, distance, and cluster capacity in the clustering process. The algorithm uses three parameters $\alpha$, $\beta$, $\gamma$ to weight these factors and optimize cluster formation.
\cite{mahyastuty2018wireless} evaluates the performance of WSN using the LEACH routing protocol with a HAP as a base station. The study analyzes energy consumption, dead node count, and packet delivery for different HAP altitudes, demonstrating that higher altitudes improve network performance.
Clustering algorithms help reduce energy consumption at the device level while optimizing the overall network lifespan.

HAPs also enable dynamic reconfiguration of network topologies in response to real-time changes. In delay-sensitive applications, for example, HAPs can reallocate bandwidth or switch relay techniques to preserve quality of service. This flexibility is absolutely essential in hybrid terrestrial-aerial networks for load balancing, congestion avoidance, and packet loss minimization.
For instance, in the study~\cite{zhu2021optimal}, the authors propose a two-stage joint resource allocation and HAP deployment solution to minimize the power consumption in space-air-ground IoT networks.To find a near-optimal solution with minimal computational complexity, the suggested algorithm splits the problem into two smaller problems. Another area of growing relevance is the use of HAPs in LEO satellite-assisted networks. HAPs reduce latency variations, so improving connectivity in crowded or mobile environments by bridging the gap between satellite links and terrestrial users. These designs show promise for scalable worldwide IoT systems especially in future 6G installations.
\cite{Benaya2025} offers a non-terrestrial Integrated Sensing and Communication (ISAC) framework utilizing HAPs and UAVs to enhance coverage, security, and computing capabilities. The framework jointly optimizes beamforming and UAV trajectory to maximize communication efficiency while considering radar performance, security, offloading, and power constraints.

HAP-based infrastructures have a number of drawbacks despite their benefits. These include atmospheric fluctuations, payload limitations, and the requirement for exact control and placement. Creating power-conscious routeing protocols, lightweight transceivers, and AI-driven optimization frameworks suited to high-altitude dynamics are all necessary to overcome these constraints.
In conclusion, HAPs are becoming essential parts of wireless networks of the future. They are perfect for advanced industrial deployments and developing regions because of their capacity to expand coverage, facilitate massive IoT connectivity, and adjust to changing environmental conditions. As these platforms evolve, integrating machine intelligence and advanced communication protocols will be key to unlocking their full potential.

\subsubsection{Advanced System Integration and Future Networks}

6G does not focus solely on simple increases in data speed and reductions in latency. The 6G approach also aims for a fundamental change to move beyond communication networks. Flexibility and real-time adaptability are required to enable such fundamental changes. Network architectures must be addressed for these changes to be implemented. 
In the context of heterogeneous networks, HAPs have become an alternative solution for diversifying network architectures. The hybrid structures of HAPs help them become flexible nodes that can meet different network needs. They can thus guarantee communication in regions with geographical, infrastructure, or environmental differences.
The use of HAPs with Terahertz (THz) communication, machine learning-based optimization algorithms, and multi-dimensional resource scheduling is important for NTNs.  Hybrid systems can dynamically reconfigure themselves to changing user needs, environmental conditions, or application-specific requirements. XR experiences, massive IoT environments, real-time industrial automation, and mobile scenarios such as URLLC are examples of use cases. The intelligent optimization of HAP deployment strategies and resource management with 3D topology-aware approaches have delivered critical innovations. 3D models provide robust service coverage by accounting for HAP altitudes and mobility models.
Recent studies underline joint optimization models addressing several interdependent variables, including HAP placement, directional beamforming, user-device association, and dynamic spectrum allocation at once. These models use deep optimization and reinforcement learning to adjust in real-time to dense, interference-prone IoT environments, where device density, traffic heterogeneity, and latency sensitivity vary significantly over time and space.
Such systems can significantly improve spectral and energy efficiency by explicitly modeling the interactions among ground terminal distribution, atmospheric conditions, high altitude, and multi-beam interference zones. This increases the general network throughput and quality of experience (QoE) for end users and extends the operational lifetime of aerial nodes (especially solar-powered platforms). The convergence of HAPs with AI-native control planes, semantic communication, and real-time orchestration layers as 6G standards crystallize is poised to redefine the basis of global connectivity, autonomous, and resilient NTN architectures.

\cite{9552535} proposes a joint 3D-location planning and resource allocation scheme for XAPS-enabled C-NOMA in 6G heterogeneous IoT networks to maximize the overall system uplink achievable sum rate. A two-stage solution is proposed to address the MINLP problem, involving a many-to-one spectrum matching game and iterative UAV location optimization. The increasing complexity of these networks also demands distributed intelligence, where learning-based agents continuously adapt to changes in user behavior, channel states, and mobility patterns. Edge computing plays a vital role here, and HAPs are well-positioned to function as aerial edge nodes capable of executing real-time analytics, caching, and decision-making functions. The study in \cite{ke2021edge} introduces an edge computing framework supported by HAP networks to tackle the challenges posed by large-scale IoT connectivity. It presents an aerial, cell-free mMIMO architecture enabled by HAPs, along with a grant-free access mechanism to ensure low latency and high efficiency for IoT connections. Similarly, \cite{yang2020high} proposes a Markov chain-based channel model tailored for HAPS communications to effectively support massive IoT deployments.

Furthermore, hybrid network architectures involving LEO satellites, UAVs, and terrestrial nodes benefit from HAPs as intelligent intermediaries. These SAGIN-style systems dynamically coordinate traffic flow, aggregate sensing data, and maximize vertical handovers employing HAPs. DRL or federated optimization techniques add even more to network resilience and performance. \cite{ei2023joint} proposes a joint HAP-LEO association and power allocation scheme to maximize data transmission and minimize energy consumption in HAP-LEO-assisted IoT networks, considering QoS requirements, HAP power budget, and LEO satellite constraints. The problem is solved by optimizing the HAP-LEO association using GUROBI and power allocation using the whale optimization method. \cite{naeem2022novel} suggests a novel frame design for ISAC in NTN by exploiting the waiting period of a pulsed radar for interference-free communication and sensing. The proposed design overcomes latency issues in conventional time division duplexing (TDD) systems and enhances spectrum efficiency. \cite{Alqasir2024} proposes an energy-efficient optimization framework for user association and power allocation in a multilayer NTN architecture consisting of HAPs, UAVs, and Ground Base Stations (GBSs). The approach minimizes total power consumption while ensuring QoS requirements for users in rural and disaster-affected areas.

While these integrations promise great advances, challenges remain. Power constraints, platform mobility and security protocols must be addressed to ensure stable and reliable operations. Furthermore, interoperability between air and ground components must be designed with attention to latency, synchronization and routing adaptability.

\begin{table*}[!ht]
\centering
\scriptsize
\renewcommand{\arraystretch}{1.05}
\caption{Comparison of Advanced Integrated Sensing studies.
           TOS = type of sensing;
           HT = heterogeneous multi‑tier design;
           EE = includes an energy‑efficiency metric or optimization
           }
\resizebox{\textwidth}{!}{%
\begin{tabular}{|c|c|c|c|c|c|c|c|}
\hline
\textbf{Study} & \textbf{Year} & \textbf{TOS} & \textbf{Sensor Type} & \textbf{Coverage Area} & \textbf{HT} & \textbf{EE}  \\
\hline
\cite{9911442} & 2022 & Environmental & LoRa & Urban & \checkmark & \checkmark \\
\hline
\cite{almarhabi2021lora} & 2021 & Environmental,Agricultural   & LoRa & Urban,rural & \checkmark & \cxmark \\
\hline
\cite{uyanik2022investigation} & 2022 & Environmental & LoRa& Rural-wide & \checkmark & \checkmark  \\
\hline
\cite{andreadis2022lowpower} & 2022 & Environmental,Agricultural  & LoRa & Rural & \checkmark & \checkmark \\
\hline
\cite{kusuma2019design} & 2019 & Environmental & WS nodes  & Urban & \checkmark & \checkmark \\
\hline
\cite{adha2019design} & 2019 & Environmental & WS nodes, GPS & Urban & \checkmark & \checkmark  \\
\hline
\cite{albagory2018novel} & 2018 & Environmental,Agricultural  & WS nodes  & Rural-wide & \checkmark & \cxmark  \\
\hline
\cite{brauchle2021towards} & 2021 & Marine & Camera & Marine-wide & \checkmark & \cxmark \\
\hline
\cite{li2023high} & 2023 & Marine & IoT sensors & Marine-wide & \checkmark & \cxmark \\
\hline
\cite{Rees2020} & 2020 & Environmental & Camera & Global & \cxmark & \cxmark  \\
\hline
\cite{zheng2023analysis} & 2023 &  Navigation, positioning & GNSS & Urban & \checkmark & \cxmark \\
\hline
\cite{wang2024covert} & 2024 & Environmental & WS nodes & Urban & \checkmark & \cxmark \\
\hline
\cite{Zhang2025AoI} & 2025 & Environmental & IoT sensors & Global,urban,rural & \checkmark & \checkmark \\
\hline
\cite{10318836} & 2023 & Environmental& IoT sensors & Rural & \checkmark & \cxmark  \\
\hline
\cite{sibiya2019reliable} & 2019 & Marine &Marine sensing devices & Marine, offshore&  \cxmark & \cxmark \\
\hline
\cite{xiao2019lstm} & 2019 & Multiple & IoT sensors & Rural-wide & \checkmark & \cxmark  \\
\hline
\cite{wang2022performance} & 2022 & Multiple & LPWAN & Global-rural & \checkmark & \checkmark  \\
\hline
\cite{mahyastuty2020clustering} & 2020 & Environmental & IoT sensors & Urban & \checkmark & \cxmark  \\
\hline
\cite{gharib2024high} & 2024 & Environmental & WS nodes & Global & \checkmark & \checkmark \\
\hline
\cite{pangestu2021parameter} & 2021 & Environmental & WS nodes  & Rural-wide & \checkmark & \checkmark \\
\hline
\cite{mahyastuty2018wireless} & 2018 & Environmental & WS nodes & Rural-wide & \checkmark & \checkmark \\
\hline
\cite{zhu2021optimal} & 2021 & Environmental & IoT sensors & Global & \checkmark & \checkmark \\
\hline
\cite{Benaya2025} & 2025 & Environmental,surveillance & Radar & Urban-wide & \checkmark & \checkmark \\
\hline
\cite{9552535} & 2021 & Environmental & IoT sensors & Global & \checkmark & \checkmark  \\
\hline
\cite{ke2021edge} & 2021 & Environmental  &  WS nodes & Global & \checkmark & \checkmark \\
\hline
\cite{yang2020high} & 2020 & Environmental & WS nodes & Urban, wide& \checkmark & \checkmark \\
\hline
\cite{ei2023joint} & 2023 & Environmental & IoT sensors & Global & \checkmark & \checkmark \\
\hline
\cite{naeem2022novel} & 2022 & Environmental, Traffic  & Radar & Urban& \checkmark & \checkmark \\
\hline
\end{tabular}}

\label{tab:is}
\end{table*}

\subsection{Advanced Aerial Computing} \label{section3}

\begin{figure*}
    \centering
    \includegraphics[width=0.99\linewidth]{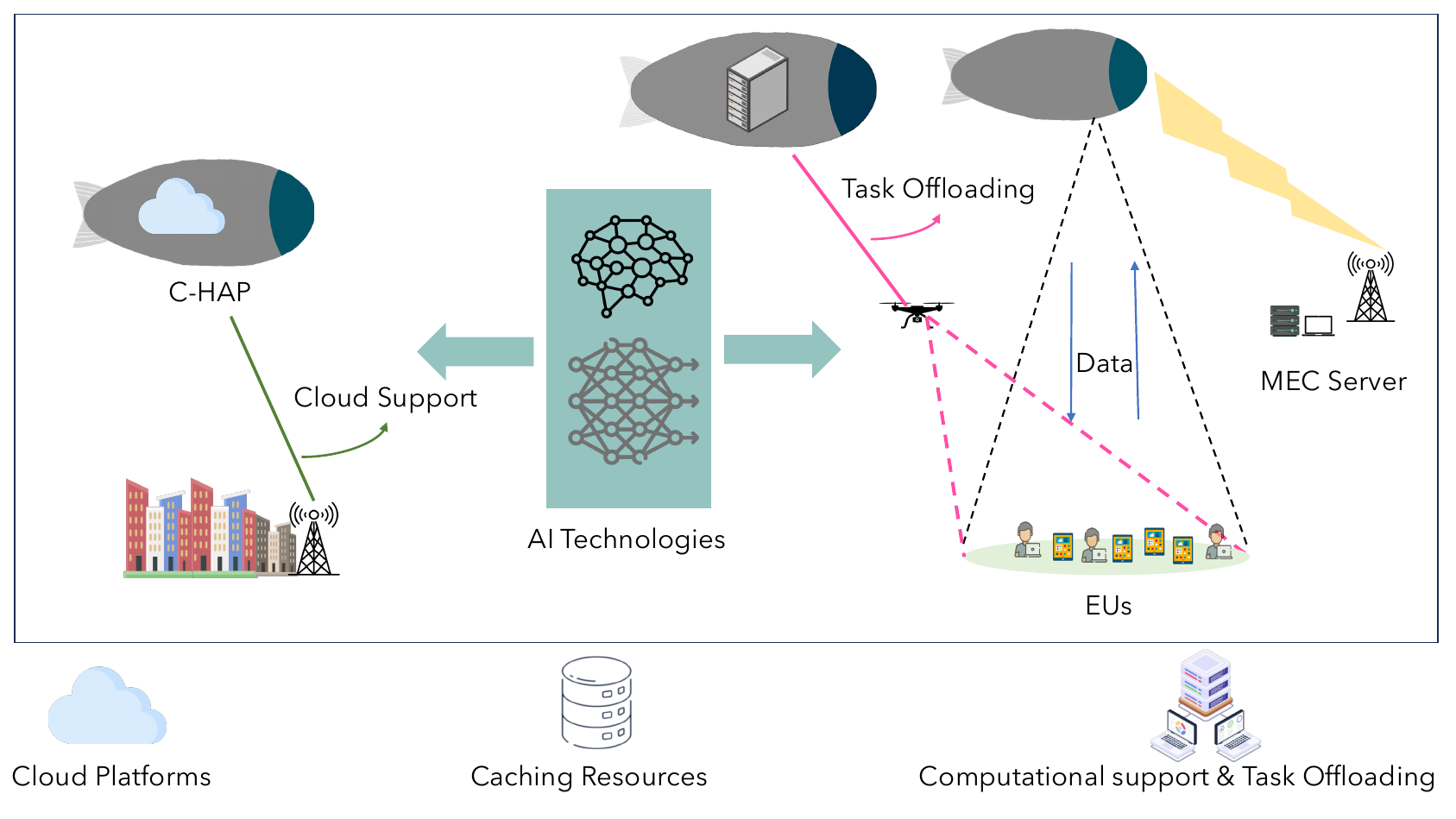}
    \caption{Computing Applications Utilizing HAPs}
    \label{fig:section4}
\end{figure*}

With HAPs essential in extending computational capabilities outside traditional terrestrial and satellite systems, advanced aerial computing has become a pillar of next-generation communication networks. Operating in the stratosphere, HAPs offer a flexible and reasonably priced answer for various uses, including real-time analytics, fast data processing, and autonomous decision-making in underdeveloped and remote locations. These airborne computing platforms enable seamless connectivity and the best use of resources over large geographical areas by bridging the gap between cloud-based processing and edge intelligence as shown in Figure \ref{fig:section4}.

Two main architectures, Edge HAP and Cloud HAP, have surfaced in advanced aerial computing. Using large-scale processing capabilities to support demanding workloads, including artificial intelligence-driven analytics, extensive dataset simulations, and cloud-assisted decision-making, Cloud HAP centers computational resources in high-powered aerial data centers. Although this centralized approach guarantees excellent scalability and efficiency, it sometimes runs across real-time responsiveness difficulties, bandwidth restrictions, and latency. Edge HAP, on the other hand, distributes processing capability nearer the data sources and end users, so decentralizing computational capacity. Edge HAP greatly improves real-time data processing by directly implementing edge intelligence on HAPs or working with lower-altitude platforms such as UAVs, thus lowering latency and enabling time-sensitive uses, including autonomous navigation, emergency response, and vehicle edge computing.

Analyzing their advantages, difficulties, and possible uses, the section investigates the basic variations between these aerial computing paradigms. Examining essential research contributions that have advanced state-of-the-art Cloud HAP and Edge HAP technologies, it also highlights creative approaches in resource management, artificial intelligence-driven optimization, and flawless integration with terrestrial and satellite networks. This framework has allowed two main architectures to be investigated. While Edge HAP uses distributed computing to improve real-time decision-making, Cloud HAP concentrates computational resources in high-powered data centers onboard the HAP. The section investigates pertinent research contributions in every field and the basic variations between these architectures.

\subsubsection{Cloud}

Cloud computing has become a critical technology offering scalable and on-demand processing capability for a broad spectrum of connected devices as demand for computational resources rises. These computational features guarantee flawless accessibility and effective operations for individuals, businesses, and large-scale companies by covering vital resources, including apps, networking, storage, data processing, and other essential services.

Three main paradigms help to define cloud computing services: Infrastructure as a Service (IaaS), Platform as a Service (PaaS), and Software as a Service (SaaS)~\cite{ahmad2021fault}. By providing virtualized computing infrastructure (including virtual machines, storage, and networking) IaaS helps customers to deploy and control their applications free from the demand for actual hardware. Operating systems, development tools, and middleware offered by PaaS create a development and deployment environment free from infrastructure maintenance so developers may concentrate on application building. Utilizing a subscription, SaaS provides software programs via the Internet, therefore removing the need for local installation and maintenance and allowing access to programs from anywhere with an Internet connection.

Cloud computing has fundamentally changed several fields of communication and networking technology, given the outstanding capabilities and ongoing improvements in cloud services. Modern computing paradigms, including edge computing, fog computing, and IoT, now include it as necessary since it allows low-latency data processing and real-time analytics. Furthermore, cloud technologies help businesses to use intelligent decision-making, automation, and predictive analytics to implement AI and ML models.

As cloud computing develops hybrid and multi-cloud approaches, where companies combining public and private cloud infrastructures to maximize cost, security, and performance, have also evolved. Cloud computing is also a must-have technology in the digital age since cloud-based networking solutions have improved the efficiency of data transmission, storage, and security in vast-scale distributed environments.

Further developments in security mechanisms, energy-efficient data centers, and quantum computing integration are expected to shape the next generation of cloud computing, reinforcing its basic importance as a fundamental technology for future innovations as businesses and people depend on cloud technologies.

However, due to growing demand and variety of uses, cloud computing technologies face some challenges, such as scalability and QoS needs~\cite{mershad2021cloud}. HAPs, with their wide range of coverage and capabilities, can offer alternative solutions to many of the challenges facing cloud technologies. On the integrated cloud infrastructure, Cloud HAP uses centralized computing resources to create a highly efficient airborne data center that supports a wide spectrum of advanced applications. High-performance servers and AI-driven processing units let Cloud HAP handle complex computations, enable machine learning model training, and run large-scale simulations otherwise limited by terrestrial networks. This capacity is beneficial when access to conventional cloud infrastructure is restricted, as in remote areas, disaster-torn regions, and maritime or aerial operations.

One critical research contribution in this domain is the development of blockchain-based security frameworks to safeguard cloud-enabled HAP networks. As proposed in recent studies, blockchain technology enhances data integrity and protects cloud transactions from cyberattacks. The study ~\cite{mershad2023blockchain} proposes an IoT cloud hap (C-HAPS) based real-time monitoring system with wireless sensor nodes. The system performs on a Sensor-as-a-Service paradigm, where HAP stations collect sensor data and make it available to cloud users. It mainly addresses infrastructure monitoring (EIM) and safe surroundings. The method primarily used blockchain for its inherent benefits, which are crucial for safeguarding cloud-enabled HAPS systems against cyberattacks: data immutability and distributed consensus procedures.

Moreover, fair access to 6G cloud services using HAPs has been considered a way to increase connectivity in areas with poor infrastructure. Studies have indicated that including HAPs in hybrid air-ground networks maximizes resource allocation and offers consistent connectivity to urban and rural sites. Convexification methods among advanced optimization algorithms have been used to optimize network throughput while lowering interference in cloud-enabled HAP systems.~\cite{alghamdi2024equitable} introduces C-HAPS architecture to optimize connectivity for hybrid air-ground networks. They suggest a practical cloud-enabled approach combining aerial platforms with terrestrial base stations to better connect underprivileged and largely populated regions. Their method optimizes beamforming vectors and user scheduling strategies using convexification approaches, including fractional programming and sparse beamforming. The findings show how well the system could offer fair service coverage and notable throughput and justice enhancements over terrestrial-only structures. Another study \cite{monzon2022high} integrates LoRaWAN gateways, HAPs, and cloud services. The system aims to acquire IoT data from isolated rural regions lacking conventional network coverage and process it using Amazon Web Services (AWS). To guarantee the smooth transmission of 
IoT data gathered from ground sensors are sent to the cloud via a 5G core network. HAPs bridge the communication gap between distant rural areas and cloud-based processing systems. Through simulations, they confirmed HAPS' capability to bridge connectivity gaps efficiently and economically, especially in remote agricultural IoT applications.

In contrast to cloud-enabled HAPs, which allow for centralized computing, edge-enabled HAPs (E-HAPs) distribute computational resources closer to end users, allowing for real-time processing. Applications, including low-latency data processing, vehicular edge computing (VEC), IoT networks, and emergency response systems depend critically on edge HAPs. One significant advancement in this field is HAP-assisted VEC for rural areas, where real-time task offloading is critical for optimizing autonomous vehicle operations. They investigated HAPs-based edge computing solutions supporting real-time data processing requirements of ground vehicles in remote areas. The study maximized computational load distribution between onboard processing and offloading activities to HAPs by modeling the network as a queuing system, significantly lowering latency and increasing general processing efficiency. Millimeter-wave communication improved vehicle network responsiveness even further and satisfied the latency requirements for autonomous driving conditions.

HAP-assisted VEC for rural areas is one significant development in this field since real-time task offloading is essential to maximize autonomous vehicle operations. \cite{cao2021edge} presents an edge-cloud resource scheduling system based on SDN for the Io in SAGIN environments. Using network function virtualization (NFV) and SDN, their solution efficiently coordinated satellite, aerial, and terrestrial resources, thus improving the general performance of vehicle networks, particularly in latency and resource optimization. Moreover, priority-aware task offloading in edge-HAP networks has been introduced as an innovative approach to improving QoS in mission-critical applications. Moreover, priority-aware task offloading in edge-HAP networks has been introduced as an innovative approach to enhancing QoS in mission-critical applications. Researchers have employed deep reinforcement learning (DRL) techniques to allocate computational resources based on task urgency dynamically. Likewise highlighting task prioritizing in satellite- and HAPs-assisted contexts, \cite{dai2023priority} concentrated on cloud and mobile edge computing (MEC) within SAGIN. They proposed a priority-aware deep reinforcement learning model to maximize task offloading and resource allocation. This method considerably lowered computational delay and improved network efficiency, so meeting QoS needs for different IoT applications with varying degrees of priority.

Collectively, these studies underline the growing significance of hybrid aerial-terrestrial systems, leveraging HAPS platforms integrated with cloud computing and advanced communication techniques. From rural IoT applications to high-capacity urban networks, these developments offer enhanced connection, best use of resources, strong security, and higher QoS across many demanding networking scenarios.
 
\begin{table*}[!htbp]
\centering
\scriptsize
\renewcommand{\arraystretch}{1.05}
\caption{Comparison of Advanced Aerial  Computing studies.
           CP = computing paradigm;
           TO = task offloading;
           EE = includes an energy‑efficiency metric or optimization
           }
\resizebox{\textwidth}{!}{%
\begin{tabular}{|l|c|p{7cm}|c|c|c|c|c|}
\hline
\textbf{Study} & \textbf{Year} & \textbf{Contribution} & \textbf{CP} & \textbf{TO} & \textbf{AI/ML} & \textbf{Caching} & \textbf{EE} \\
\hline
\cite{mershad2023blockchain} & 2023 & Blockchain-secured C-HAPS system & Cloud-based & \cmark & \cxmark & \cxmark & \cmark \\
\hline
\cite{alghamdi2024equitable} & 2024 & Equitable 6G connectivity via C-HAPS  & Cloud-based & \cmark & \cmark & \cxmark & \cxmark \\
\hline
\cite{monzon2022high} & 2023 & Smart rural IoT using HAPS and 5G cloud integration & Cloud-based & \cmark & \cxmark & \cxmark & \cxmark \\
\hline
\cite{cao2021edge} & 2022 & Multi-layer SDN control with many-objective optimization in SAGIN-IoV. & Hybrid & \cmark & \cxmark & \cxmark & \cxmark \\
\hline
\cite{dai2023priority} & 2024 & DDPG-based priority-aware offloading and resource allocation  & Hybrid & \cmark & \cmark & \cxmark & \cxmark \\
\hline
\cite{umar2024computation} & 2024 & Power and resource optimization method  with NOMA-ME& Edge-based & \cmark       & \cxmark  & \cxmark    & \cmark        \\
\hline
\cite{mei2022energy} &  2024 & Heuristic optimization algorithm for  HAP-satellite edge computing systems & Edge-based           & \cmark & \xmark  & \xmark    & \cmark \\
\hline

\cite{ren2022caching} &   2024 &  Computational offloading and caching optimization framework  & Edge-based & \cmark       & \xmark  & \cmark    & \cmark        \\
\hline
\cite{waqar2022computation} &   2024 & Decentralized DRL framework to optimize computation offloading & Edge-based & \cmark       & \cmark  & \xmark    & \cmark        \\
\hline
\cite{gong2022computation} &   2024 & Lyapunov-guided MARL framework for computation offloading & Edge-based  & \cmark & \cmark  & \xmark    & \cmark        \\
\hline
\cite{wang2024computation} &   2024 & MADRL algorithmfor computation offloading decisions & Edge-based           & \cmark       & \cmark  & \xmark    & \cmark        \\
\hline
\cite{masood2021content} &   2024 & Hierarchical FL-based caching framework & Edge-based & \cmark       & \cmark  & \cmark    & \xmark        \\
\hline
\cite{kang2023cooperative} &   2024 & MAPPO-based MARL framework for task offloading & Edge-based           & \cmark       & \cmark  & \xmark    & \cmark        \\
\hline
\cite{li2024dynamic} &   2024 & Lyapunov-based dynamic energy-efficient algorithm & Edge-based  & \cmark       & \xmark  & \xmark    & \cmark        \\
\hline
\cite{li2024dynamic} &   2024 & Dynamic weighting-based optimization framework & Edge-based           & \cmark       & \xmark  & \xmark    & \cmark        \\
\hline
\cite{li2024dynamic} &   2024 & Hierarchical computing framework using matching-game theory & Edge-based  & \cmark       & \xmark  & \xmark    & \cmark        \\
\hline
\cite{nguyen2024intelligent} &   2024 & a MADDPG-based intelligent aerial edge computing framework                           & Edge-based           & \cmark       & \cmark  & \xmark    & \xmark        \\
\hline
\cite{lakew2023intelligent} &   2024 & MADRL approach for joint optimization of IoTD association, partial offloading, and resource allocation                                   & Edge-based           & \cmark       & \cmark  & \xmark    & \cmark        \\
\hline
\cite{li2024joint} &   2024 & Framework for minimizing task offloading delay                                             & Edge-based           & \cmark       & \xmark  & \xmark    & \cmark        \\
\hline
\cite{Lu2024} &   2024 & MADRL-based joint dynamic role switching and cooperative offloading algorithm         & Edge-based           & \cmark       & \xmark  & \xmark    & \cmark        \\
\hline
\cite{yuan2023joint} &   2024 & Joint optimization of content caching, offloading, and server selection                        & Edge-based           & \cmark       & \xmark  & \cmark    & \cmark        \\
\hline
\cite{9999999} &   2024 &  Hierarchical aerial computing platform with a joint offloading decision & Edge-based           & \cmark       & \xmark  & \xmark    & \cmark        \\
\hline
\cite{sun2024joint} &   2024 & Optimization for UAV trajectory, caching, and task offloading                                              & Edge-based           & \cmark       & \xmark  & \cmark    & \cmark        \\
\hline
\cite{Do2024} &   2024 &  DDPG-based algorithm for QoE optimization in green AEC networks     & Edge-based           & \cmark       & \xmark  & \xmark    & \cmark        \\
\hline
\cite{truong2022mec} &   2024 &  DDPG-based algorithm  for optimizing task offloading                                   & Edge-based           & \cmark       & \xmark  & \xmark    & \cmark        \\
\hline
\cite{Tang2025MTrain} &   2024 & MTrain scheduling strategy leveraging sub-operation parallelism           & Edge-based           & \cmark       & \xmark  & \xmark    & \cmark        \\
\hline
\cite{Wu2025} &   2024 &SAGSIN computation offloading architecture                & Edge-based           & \cmark       & \xmark  & \xmark    & \cmark        \\
\hline
\cite{nguyen2022multiagent} &   2024 & MADRL approach for task offloading  with NOMA                                                                                               & Edge-based           & \cmark       & \xmark  & \xmark    & \cmark        \\
\hline
\cite{Fettes2024} &   2024 & HAPS-enabled architecture to enhance optical data transfer & Edge-based           & \cmark       & \xmark  & \xmark    & \xmark        \\
\hline
\cite{that2024optimizing} &   2024 & DDPG-based framework for jointly optimize UAV placement and offloading decisions                        & Edge-based           & \cmark       & \xmark  & \xmark    & \cmark        \\
\hline
\cite{dai2023priorityaware} &   2024 &  Priority-aware task offloading and resource allocation scheme                         & Edge-based           & \cmark       & \cmark  & \xmark    & \cmark        \\
\hline
\cite{1570914282} &   2024 & MEC framework utilizing NOMA in a multi-layer HAP and UAV network                             & Edge-based           & \cmark       & \xmark  & \xmark    & \cmark        \\
\hline
\cite{Peng2024TaskOffloading} &   2024 & Iterative maximum flow algorith for task offloading                                          & Edge-based           & \cmark       & \xmark  & \xmark    & \xmark        \\
\hline
\cite{kawamoto2024traffic} &   2024 &  Dynamic scheduling strategy for resource control in MEC-assisted satellite communication                          & Edge-based           & \cmark       & \cmark  & \xmark    & \cmark        \\
\hline
\cite{Li2025TwoHop} &   2024 & DRL-based method for joint task offloading, resource allocation                                      & Edge-based           & \cmark       & \xmark  & \xmark    & \xmark        \\
\hline
\cite{gao2024intelligent} &   2024 & DRL algorithm for task offloading and resource allocation                            & Edge-based           & \cmark       & \cmark  & \xmark    & \xmark        \\
\hline
\cite{Ahsan2025Computational} &   2024 & DRL-based offloading and time allocation policy                                                    & Edge-based           & \cmark       & \cmark  & \xmark    & \cmark        \\
\hline
\cite{chen2024online} & 2024& Online energy-efficient dynamic offloading algorithm                                                                      & Edge-based           & \cmark       & \xmark  & \xmark    & \cmark        \\
\hline
\end{tabular}}

\label{tab:aac}
\end{table*}

\subsubsection{Edge} 

As an alternative to cloud computing, computational capabilities are moved to edge networks. Unlike the cloud, this offers end users a more beneficial experience in terms of QoS. 
HAPS integrated with UAVs transforms MEC networks by enabling flexible, scalable, and location-independent task processing. This paradigm shift addresses the computational limitations of IoT and edge devices by allowing partial or complete task offloading to aerial servers, where computation and storage resources are dynamically allocated. In such networks, resource allocation becomes a multidimensional optimization problem involving bandwidth, computing power, and energy budgets.

The offloading decision involves choosing between UAVs for immediate, low-latency tasks or HAPS for high-throughput processing, often across multi-hop or hierarchical configurations. Recent literature adopts deep learning and heuristic approaches for decision-making, considering energy consumption and delay constraints. Advanced scheduling strategies exploit the cooperation between UAVs (as mobile edge servers or relays) and HAPS (as stable computing nodes) to optimize system performance jointly. 
\cite{kang2023cooperative} investigates a hierarchical aerial computing framework integrating HAPs and multiple UAVs to provide efficient computation services for ground devices (GDs), particularly in scenarios like disaster recovery where terrestrial infrastructure may be damaged or unavailable. The authors propose a cooperative resource allocation and task offloading scheme to address the challenges of limited UAV resources, heterogeneous QoS requirements of GDs, and the long transmission delays associated with direct HAP-GD communication. The scheme is formulated as a partially observable Markov decision process (POMDP), which jointly optimizes UAV resource allocation (including spectrum, computing, and caching resources) and the ratio of tasks offloaded to HAPs. A multi-agent proximal policy optimization (MAPPO)-based reinforcement learning algorithm is designed with centralized training and decentralized execution to ensure scalability and adaptability. 
\cite{Ahsan2025Computational} proposes an offloading and time allocation policy using Mobile Device Cloudlet (MDC) and MEC (OTPMDC) to improve energy efficiency and reduce latency in wireless-powered MEC systems. It trains a deep learning-based decision-making algorithm to choose an optimal set of applications based on the energy in the User Equipage (UE). The simulation results show that the performance of UEs is improved.

For UAV swarm-enabled environments, the study \cite{Lu2024} offers a joint dynamic role switch scheme and cooperative offloading algorithm (MARSCO) for UAV swarm-enabled edge computing to minimize the total system latency and energy consumption, subject to constraints on battery capacity and execution latency. The proposed algorithm utilizes multi-agent deep reinforcement learning (MADRL) to optimize two sub-problems iteratively: dynamic role switching and cooperative computing offloading.
 A hierarchical aerial computing platform~\cite{9999999} leveraging UAVs and HAP to meet the computation demands and latency requirements of various IoT applications for GUs. The study introduces a joint offloading decision, user association, and resource allocation (JOUR) scheme, utilizing binary offloading from GUs to UAVs and partial offloading from UAVs to HAP, minimizing energy consumption and latency while maximizing load balancing.
 UAV trajectory planning is also crucial for task offloading MEC networks. \cite{dai2023priorityaware}, the authors introduce a priority-aware edge computing framework and trajectory optimization approach tailored for SAGIN comprising a large number of IoT devices. The proposed model aims to address challenges in large-scale access, task offloading, and resource management while delivering computing services with differentiated priority levels to IoT devices in remote regions. To enhance system performance, the study employs a priority-aware TDMA-based MAC protocol (PTDMA) and a DDPG-based scheme named PDOA for task offloading and resource allocation.
Similarly, \cite{Li2025TwoHop} focuses on minimizing task offloading delays in air–ground integrated MEC systems. The study jointly optimizes partial offloading decisions, resource allocation, and UAV trajectory planning. To address the complexity of the problem, it is decomposed into two subproblems and solved using DRL-based methods, specifically, the MADDPG-IPER and NV-IPPO algorithms.

These systems are advantageous in environments without terrestrial infrastructure, such as oceans, deserts, or disaster zones. Moreover, intelligent resource slicing and real-time optimization support latency-sensitive and bandwidth-hungry applications like augmented reality, autonomous vehicles, and smart agriculture. 
\cite{umar2024computation} proposes a dual-layer optimization framework for an uplink NOMA-enabled aerial-vehicular network integrated with MEC using millimeter wave (mmWave) technology, specifically addressing computation offloading through optimized resource allocation and data-aware clustering to reduce transaction latency and enhance throughput significantly.
The convergence of air-ground networks into a cohesive computational layer through HAPS-UAV collaboration marks a significant evolution in MEC research.
Also \cite{li2024joint}  addresses the challenge of minimizing task offloading delay in air-ground integrated vehicular edge computing VEC) networks. It proposes a joint multicomputation equipment selection and multidimensional resource allocation (JCESRA) algorithm, considering HAPs, UAVs, and roadside units (RSUs) to optimize resource allocation and reduce delay.

The application of AI, particularly RL, is revolutionizing the design and management of aerial edge computing systems, including HAPS and UAVs. These intelligent systems address the limitations of traditional static networks by enabling autonomous decision-making and real-time optimization in highly dynamic environments.DRL algorithms, such as DDPG, PPO, MADDPG, and MAPPO, have extensively optimized computation offloading, UAV trajectory planning, resource scheduling, and user association.
In multi-agent settings, DRL allows coordinated task offloading where UAVs and HAPSs act as agents that learn policies to minimize energy consumption, reduce latency, or maximize throughput. This learning-based optimization benefits multi-layer aerial networks where task assignment, routing, and coverage planning must adapt to unpredictable user demands and environmental conditions. Additionally, DRL-driven strategies address non-convex and large-scale problems where traditional optimization fails to converge efficiently.
Several studies in the literature offer multi-agent-based solutions. 
Study \cite{nguyen2022multiagent}  presents a hierarchical HAP-LAP network with NOMA support, where both HAPs and LAPs provide computational services for IoT devices. Simultaneously, LAPs serve as intermediaries that forward tasks to HAPs. This process frames the task offloading challenge using a partial offloading approach under QoS constraints, aiming to reduce overall energy consumption and execution delays across all devices. The problem is modeled as a POMDP and tackled using the MADDPG algorithm.

Furthermore, hierarchical reinforcement learning brings abstraction to multi-service edge systems, improving scalability and policy convergence. Priority-aware decision-making made possible by AI-based systems guarantees fairness and QoS variation among different IoT activities. These intelligent aerial networks hold promise for future 6G systems, enabling real-time adaptability and self-optimization across space-air-ground architectures. 
\cite{wang2024computation} presents a multi-agent DRL-based approach for computation offloading in an aerial hierarchical MEC system consisting of HAPs and UAVs. Recognizing limitations in traditional terrestrial edge computing networks (such as inadequate coverage, limited computational resources, and high latency) the authors propose an aerial MEC architecture where UAVs provide edge computing services to IoT devices, with HAPs offering additional computational support. They formulate the computation offloading as a long-term cost minimization problem considering task queuing mechanisms, non-divisible task constraints, processing delays, and abandonment penalties. To handle dynamic and uncertain task arrivals, the authors introduce a DRL-based algorithm enhanced by convolutional Long Short-Term Memory (ConvLSTM) networks for predicting UAV task loads and Prioritized Experience Replay (PER) for improved training efficiency and convergence stability.
Furthermore, \cite{gao2024intelligent} explores the challenges of task offloading and resource allocation in collaborative networks composed of multiple HAP drones and satellites. The study formulates an optimization problem addressing task partitioning and joint communication-computation resource management, aiming to minimize the overall task delay for IoT devices. A deep reinforcement learning-based algorithm is developed to optimize task splitting decisions.
In a related vein, \cite{lakew2023intelligent} addresses the integrated challenges of IoT device association, partial task offloading, and communication resource allocation in heterogeneous Aerial Access IoT (AAIoT) environments. The goal is to enhance service satisfaction while reducing energy consumption. To tackle this, the authors propose MADDPG-JAPORA, a multi-agent policy-gradient deep actor-critic algorithm that models the problem as a multi-agent Markov decision process (MAMDP).
Meanwhile, \cite{truong2022mec} examines MEC-supported aerial network scenarios where aerial platforms, with assistance from HAPs, deliver services to remote regions. The study introduces a deep reinforcement learning framework named HAMEC, built on the deep deterministic policy gradient (DDPG) algorithm, to optimize task completion by minimizing total cost.
Lastly, \cite{that2024optimizing} proposes a DRL-based framework for optimizing energy consumption and latency in air-ground integrated IoT networks. The framework leverages UAVs and HAPs for edge computing and dynamically adjusts network parameters using DDPG. 

As AI models evolve, integrating explainability, federated learning, and lightweight deployment strategies will be critical for expanding their practical applicability in aerial edge networks. In sum, reinforcement learning enables the creation of autonomous, resilient, and high-performing HAPS-UAV ecosystems that fulfill the computation-intensive demands of the next-generation wireless world.

Integrating caching and FL mechanisms into  HAPS-assisted networks has opened new horizons for efficient data management and low-latency service provisioning in next-generation communication systems. As the demand for content-centric services and real-time intelligence grows, HAPS-enabled caching frameworks offer a promising alternative to mitigate backhaul congestion and reduce service delay, particularly in remote or disconnected regions.

HAPS nodes, equipped with storage capabilities, can serve as intermediate cache servers between terrestrial content providers and end users. This hierarchical architecture is particularly effective in multi-UAV or multi-HAPS environments where spatial diversity is leveraged for dynamic content distribution. Joint optimization schemes are often employed to determine what content should be cached, where it should be placed, and how frequently it should be refreshed, all under constraints like bandwidth, delay, and energy consumption.
\cite{ren2022caching} suggests a novel three-layer computational offloading framework integrating HAPS, RSUs, and connected autonomous vehicles (CAVs) to enhance the performance of ITS. Recognizing the constraints on energy, latency, and computational capacity at terrestrial edges, the authors propose using HAPS as a stratospheric computing platform due to its extensive coverage, strong computational power, and sustainable energy supply. To minimize system delays, the framework optimizes caching strategies at terrestrial edges, task offloading decisions, and resource allocation. 
Furthermore, \cite{yuan2023joint} examines the cache-enabled high-low-altitude-platforms integrated network (CHLIN) and proposes a genetic-algorithm-embedded distributed alternating direction method of multipliers (GA-DADMM) to optimize content caching policies, offloading decisions, and HAP/LAP server selection, aiming to reduce content transmission delay while meeting user quality of service demands. 

Moreover, FL has been proposed as a distributed AI solution for model training across HAPS-assisted networks. In FL-based frameworks, edge devices collaboratively learn a shared model without exposing raw data, preserving privacy while ensuring intelligence at the edge. The hierarchical nature of HAPS-UAV systems enables a multi-tiered federated structure, where UAVs collect and preprocess data and HAPS nodes act as aggregation points for local models.
\cite{masood2021content}  proposes an intelligent, privacy-preserving content caching framework utilizing hierarchical FL(HFL) in HAP-assisted multi-UAV networks. Given the increasing demand for bandwidth-intensive applications, particularly video content, and the corresponding challenges posed by limited backhaul resources, the authors address the need for accurate prediction and caching of popular content at UAVs. The proposed HFL-based framework involves mobile UEs, UAVs, and HAP in a hierarchical structure. UEs train local deep learning models based on their content preferences and usage patterns and periodically upload model weights to nearby UAVs, which aggregate these local models using Federated Averaging (FAVG). UAVs then transmit these partially aggregated models to the HAP to perform a global aggregation and it significantly improves prediction accuracy and reduces content access latency without compromising user privacy.
On the other hand, \cite{Tang2025MTrain} suggests MTrain, a novel multi-accelerator training scheduling strategy that transfers the training process into a multi-branch workflow, enabling independent sub-operations to be executed on different training accelerators in parallel for better utilization and reduced communication overhead. Experimental results demonstrate efficient CNN training on heterogeneous FPGA-based edge servers with significant speedup compared to state-of-the-art methods.

This synergy between caching and learning enables adaptive content delivery, personalized services, and proactive network management. Additionally, intelligent caching decisions based on learned user behavior patterns can significantly improve the content-hit ratio and reduce the computational burden. As research continues, the focus is shifting toward integrating FL with content-aware caching and real-time UAV coordination and  aiming for a fully autonomous and intelligent aerial information ecosystem.




Key research challenges in NTN component inherent mobility, energy constraints, and environmental variability are task offloading, data routing, and trajectory optimization. Often supported by heuristic optimization or reinforcement learning, intelligent routing protocols are needed to control dynamic topologies and reduce latency so, guaranteeing service continuity.

For mission-critical applications in intelligent transportation, remote sensing, and marine monitoring, integrating MEC into NTNs also enables distributed data processing closer to end users. Predictive models for traffic demand and resource control are essential for managing limited bandwidth and dynamic link availability, especially in satellite-HAPS coordination~\cite{saeed2021pointtopoint}. 
\cite{kawamoto2024traffic} offers a dynamic scheduling strategy for resource managment in HAPS-mounted MEC-assisted satellite communication architectures, aiming to reduce resource wastage by predicting traffic demand and allocating resources efficiently. The proposed approach demonstrates high accuracy in traffic prediction and resource allocation, improving resource utilization.
\cite{Wu2025} proposes a novel multi-HAP-assisted SAGSIN computation offloading architecture to utilize cross-regional idle resources fully. It formulates a joint optimization problem to minimize the delay, which is solved by a graph theory-based iterative optimization scheme. 
In addition to that, \cite{Peng2024TaskOffloading} investigates task offloading in a terrestrial-support-free multi-layer multi-access edge computing (AC-TMMEC) system for mMTC beyond 5G, identifies data routing and trajectory planning as key challenges and proposes an iterative maximum flow algorithm (IMFA) to solve source-destination pairing and data routing jointly.


Deep integration of system architecture, energy efficiency techniques, and intelligent resource management is necessary in design and optimization of HAPS-based networks. These platforms, which offer persistent connectivity, computing services, and wide-area coverage free from the limitations of terrestrial installations, are seen as fundamental parts of next-generation communication infrastructures.
Architecturally, HAPS nodes have double purposes: they act as MEC server communication relays. This duality calls for careful balancing of computational load, energy consumption, and connectivity across very dynamic aerial topologies. Optimizing UAV/HAP location and power allocation is central to these initiatives since it influences throughput, latency, and energy consumption all around. Several studies consider energy consumption and optimization as well~\cite{li2024dynamic,2li2024dynamic,Fettes2024,gong2022computation}.
\cite{mei2022energy} address the challenge of high energy consumption faced by ground user equipment (GUE) when communicating with low-earth orbit (LEO) satellites for MEC. The authors propose a hybrid system integrating HAPs and multiple LEO satellites, called HAP-Satellites Edge Computing, to mitigate this. They develop an optimization framework aiming to minimize the weighted sum of energy consumption by jointly optimizing task offloading decisions, transmission power, and computational resource allocation. The problem is formulated as a mixed-integer nonlinear programming (MINLP) model, which is difficult to solve directly. Hence, the authors present an intelligent heuristic algorithm to solve the optimization efficiently.
\cite{waqar2022computation} offers joint computation offloading and resource allocation problems in integrated aerial-terrestrial vehicular networks enhanced by MEC. Understanding the constraints of terrestrial MEC networks in far-off locations, the authors suggest a novel network architecture using HAPs furnished with MEC servers linked with a backhaul network of LEO satellites. The main objective is to minimize the overall computation and communication overhead by optimizing offloading decisions, sub-band allocation, transmission power, and computing resource distribution.
\cite{chen2024online} investigates the computation offloading and resource allocation problem in a HAP-assisted MEC system to minimize energy consumption. It proposes an online Energy Efficient Dynamic Offloading (EEDO) algorithm that effectively reduces energy consumption while maintaining system stability.
In addition, \cite{nguyen2024intelligent} proposes an intelligent, heterogeneous aerial edge computing architecture leveraging NOMA, HAPs, and UAVs to efficiently support computationally intensive tasks of IoT devices IoTDs in challenging environments. Addressing the limitations of traditional terrestrial networks, such as limited coverage, latency issues, and vulnerability to infrastructure damage, the authors introduce a MADDPG-based RL algorithm to jointly optimize IoTD association, task offloading ratios, transmission power allocation, and computational resource allocation. 
Similar to \cite{nguyen2024intelligent}, \cite{1570914282} proposes a secure MEC multi-layer air-ground network model that employs NOMA for improved spectral efficiency and PLS for securing wireless transmission, aiming to maximize the system's minimum secure offloading sum rate by jointly optimizing task computation latency, task scheduling, and transmit power.
\cite{Do2024} suggests a green aerial edge computing (AEC) architecture where UAVs harvest energy from renewable resources. It formulates an optimization problem to maximize long-term mobile device (MD) satisfaction while ensuring sustainable UAV operation by controlling computation offloading and resource allocation. DDPG algorithm is developed to solve the problem in a dynamic network environment, integrating prioritized experience replay and weighted importance sampling techniques to improve learning performance.
Another approach, \cite{jia2022hierarchical} introduces a hierarchical computing framework using matching-game theory and heuristic algorithms to optimize computational offloading between IoT devices, UAVs, and HAPs. Understanding the constraints of IoT devices, limited computational power, limited energy resources, and strict delay requirements, the suggested approach uses UAVs to manage lightweight computing tasks because of their proximity to IoT devices. On the other hand, UAVs send heavy computing chores to HAPs, which have more computational and energy capability. The authors formulate the problem as an integer programming optimization task, aiming to maximize the total amount of IoT data successfully computed under strict delay and resource constraints.
\cite{sun2024joint}  proposes a framework for aerial MEC using HAPs and UAVs to optimize workload fairness for UAVs while minimizing the weighted processing costs for IoT devices.

The complexity and dynamism of aerial environments are among the main difficulties in HAPS-assisted MEC networks. Including HAPS with UAVs changes conventional MEC models but creates some challenges. First of all, optimal task offloading is intrinsically a multi-objective problem that includes latency, energy consumption, computational capacity, and user QoS requirements. This complexity is magnified in heterogeneous aerial networks where tasks must be distributed between fast-moving UAVs and more stable but distant HAPS nodes. Secondly, aerial networks' dynamic and decentralized nature requires robust scheduling and resource management under uncertain network conditions. Reinforcement learning and multi-agent systems have shown promise, yet they struggle with convergence and scalability in large-scale scenarios with highly dynamic task loads.
Furthermore, trajectory planning for UAVs complicates the optimization scene directly influences communication stability and energy economy. Limited energy resources, variable link quality, and long-range communication delays are additional bottlenecks between HAPS and ground devices. When caching and FL are introduced to improve content delivery and privacy, the trade-offs between model accuracy, communication overhead, and device heterogeneity must also be managed. These challenges call for cross-layer design strategies that intelligently coordinate real-time networking, computation, and energy policies.

Despite these challenges, integrating HAPS and UAVs within MEC frameworks marks a critical evolution in wireless communication and edge intelligence. Architectures offer low-latency, high-throughput, and scalable service provisioning by bringing computation closer to end users in disconnected or infrastructure-deficient areas. Strong reinforcement learning and federated learning allow autonomous, adaptive, and cooperative decision-making over the space-air-ground continuum. Furthermore, hierarchical computing systems and cache techniques help reduce response times and balance workloads. Emerging solutions, including green energy collecting, modular HAPS designs, and energy-aware QoE measurements, are expected to improve the viability and sustainability of these systems even more as research goes on. Ultimately, HAPS-assisted MEC networks will serve as key enablers of 6G and beyond, supporting mission-critical applications in areas ranging from smart agriculture and disaster response to autonomous mobility and immersive virtual environments.

\section{Challenges and Future Direction}

In non-terrestrial networking, HAPs offer a revolutionary technological development that bridges gaps left by conventional terrestrial and satellite systems. HAP networks have great potential and capacity, but they also present several major obstacles that need to be resolved if we are to guarantee efficient, consistent, and broad acceptance.

\subsection{Energy and Payload Capabilities}
Managing and optimizing few resources is one of the main difficulties HAP networks encounter. Although solar-powered systems present sustainable energy solutions, storage and effective management of this energy (especially during extended night operations or bad weather) remains a major issue. Directly affecting payload capabilities, flight duration, and general operational sustainability is the energy constraint. Dealing with this challenge calls for developments in lightweight, high-density energy storage technologies, solar cell efficiency gains, and intelligent energy management systems.

HAPs also have great difficulty optimizing payloads. Development of high-performance but lightweight payloads for advanced sensing, communication, and computational activities is especially difficult given strict weight constraints and energy budgets. Maximizing value and efficiency requires innovations in miniaturization, materials science, and integrated multifunctional payloads able of simultaneously performing communication, computing, and sensing tasks.

\subsection{Network and Communication}
Dynamic network topology management presents still another difficult problem. HAPs have operational altitude and mobility, thus they have to keep constant and dependable communication links even if the geographical and atmospheric conditions are always changing. Dynamic flight path adaptation in response to changing environmental conditions (including wind currents, weather disturbances, and user demand distribution) because dependent on real-time trajectory optimization. Maintaining continuous service also depends on building strong handover protocols for seamless connectivity between many HAPs, terrestrial base stations, and satellite links.

Further major challenges are spectrum management and regulatory limitations. Significant international and diplomatic challenges arising from the harmonization of international regulatory standards, frequency allocation and cross-border activities can prevent deployment in a timely and efficient manner. Moreover, effective spectrum use depends  on the development of dynamic frequency allocation techniques and adaptive spectrum-sharing systems.

Integration of cutting-edge technologies like RIS complicates HAP network operation and design more still. RIS technology adds more levels of complexity in terms of real-time configuration management and optimization, and it greatly improves communication dependability, spectral efficiency, and network adaptability. Implementing intelligent algorithms able of dynamically adjusting RIS settings to maintain optimal performance under fast changing conditions remains difficult and calls for great computational resources and advanced AI-driven solutions.

\subsection{Digital Twin}
Integrating Digital Twins (DTs) can greatly improve the ability of HAP networks to operate and control. Continuous real-time synchronization of physical HAPs with their digital counterparts gives operators situational knowledge a. two-way data sharing makes it possible to accurately model things like the environment, how much energy is used, how well payloads work, and how much network traffic is needed. So, network managers can change flight paths on the fly, guess what might go wrong, and plan ahead for how to divide up resources. Also, operators can quickly respond to changing working conditions and user needs.

HAP networks can be made more flexible and adaptable by adding DT technology. HAP DTs create varied operational situations, like severe weather, system failures, and cyber attacks. The outcome allows operators test and make the system more resilient early on. DTs can continuously monitor the condition of embedded systems, predict when parts will fail, and plan how to prevent failures. 

Furthermore, supported by DT-enabled HAP systems are complex computational and communication chores, especially in relation to edge computing models. HAP-based edge nodes combined with Digital Twin models help maximize computational offloading techniques, guaranteeing effective resource use and low latency. Real-time network conditions, computational needs, and energy restrictions help to clearly allocate tasks generated by IoT devices or user equipment among aerial nodes, so greatly improving general network efficiency and service quality.

Furthermore, DT systems included in HAP networks can maximize resource management and adaptive network behaviors by using sophisticated AI-driven algorithms, including ML and DRL. These intelligent algorithms ensure the best network performance under fast-changing circumstances and continuously enhance decision-making processes by means of real-time and historical operational data. When several HAPs and UAVs cooperate to provide integrated sensing, communication, and computational services in complex, multi-agent coordination scenarios, such AI-driven DT integration is especially advantageous.

Even with great promise, including DT technologies with HAP systems presents major technical and operational difficulties. Accurate DTs demand high-fidelity modeling, which calls for strong and consistent real-time data acquisition, significant computational resources, and sophisticated algorithms able to manage vast amounts of heterogeneous data. Maintaining data privacy and cybersecurity in DT-enabled systems is also vital, particularly given these systems are growing significant for sensitive uses including military operations, public safety, and disaster response.

As a result, the combination of DT technologies with HAP networks can provide greater flexibility and performance. Interdisciplinary studies offer global approaches to 6G systems through DT and HAP integration. Along with AI-based technological developments, SAGIN and DT-based technologies are playing a role in the air layer\cite{10345669}.

\section{Conclusion}

HAP deployment leads to significant progress in global communication, sensing, and networking. HAPs serve as an intermediate layer between terrestrial and satellite systems. With their capabilities, HAPs provide large-scale coverage and continuous connectivity in remote areas. Thus, communication and connectivity technologies become global. Additionally, HAPs offer a flexible and interactive communication infrastructure for agriculture, environmental monitoring, smart transportation, and disaster management.

The rapid deployment, flexible coverage area, and low-latency communication provided by HAPs enable their use in many application scenarios. The use of new technologies such as RIS, artificial intelligence, and edge computing with HAPs enhances their performance. These technologies enable HAPs to dynamically change their service offerings, control electromagnetic interference, maximize resource allocation, and increase spectral efficiency. With these developments, adaptation to changing needs and environmental conditions increases.

Additionally, the deployment of HAP networks is economically sustainable and offers cost-effective alternatives to traditional satellite-terrestrial infrastructure. Optimization efforts in energy and payload technologies make HAPs environmentally friendly.

Despite advances in HAP technologies, there is still a need for new developments related to energy constraints, payload limits, legal systems, and international frequency allocation. To achieve more widespread acceptance and global integration, these issues must be addressed. The sustainability and effectiveness of HAP networks depend on cybersecurity, interoperability across multiple platforms, and ensuring strong regulatory compliance.

Future research and development are required to maximize HAP performance, particularly in terms of improving energy efficiency, so extending computational capabilities, and so refining real-time adaptive mechanisms. By using AI  techniques, HAPs will be able to control resources independently, forecast network needs, and improve general service quality, so augmenting their capabilities.

As HAP technology develops, it will eventually become increasingly important for the development of next-generation networks, so helping to create a more linked, intelligent, and resilient worldwide communication scene. This future vision sees HAPs as indispensable tools necessary to propel global social equity and sustainable development and advance communication technologies.
\section{Acknowledgment}
This work was supported  in part by Istanbul Technical University, Department of Scientific Research Projects (ITU-BAP) under Grant
45375; and in part by Türkiye-South Korea Bilateral Call under Project TUBITAK 2523-123N821; and in part by the National Research
Foundation of Korea under Grant RS-2023-NR121113.





\bibliographystyle{cas-model2-names}

\bibliography{cas-refs}

\end{document}